\documentclass[journal]{IEEEtran}

\usepackage{graphicx}
\usepackage{amssymb}
\usepackage{amsmath}
\usepackage{cite}
\usepackage{mathrsfs}
\usepackage[usenames, dvipsnames]{color}
\usepackage{float}
\usepackage{subfigure}
\usepackage{epstopdf}
\usepackage{mathrsfs}
\usepackage{enumerate}
\renewcommand{\baselinestretch}{0.98}
\usepackage{lipsum}
\usepackage{algorithmic}

\renewcommand\IEEEkeywordsname{Keywords}
\newcommand{\ignore}[1]{}

\graphicspath{ {D:/} }

\begin{document}
\title{\LARGE \bf  Efficient, Flexible and Secure Group Key Management Protocol for Dynamic IoT Settings}
\author{Adhirath Kabra, Sumit Kumar and Gaurav S. Kasbekar}

\maketitle

{\renewcommand{\thefootnote}{} \footnotetext{A. Kabra, S. Kumar and G. S. Kasbekar are with the Department of Electrical Engineering, Indian Institute of Technology Bombay, Mumbai, India. Their email addresses are adhirathkabra@gmail.com, ism.sumit@gmail.com and gskasbekar@ee.iitb.ac.in, respectively.}}

\begin{abstract}
Many Internet of Things (IoT) scenarios require communication to and
data acquisition from multiple devices with similar functionalities.
For such scenarios, group communication in the form of multicasting and broadcasting has proven to be effective. Group Key Management (GKM) involves the handling, revocation, updation and distribution of cryptographic keys to members of various groups. Classical GKM schemes perform inefficiently in dynamic
IoT environments, which are those wherein nodes frequently leave or join a network or migrate from one group to another over time. Recently, the `GroupIt' scheme has been proposed for GKM in dynamic IoT environments. However, this scheme has several limitations such as vulnerability to collusion attacks, the use of computationally expensive asymmetric encryption and threats to the backward secrecy of the system. In this paper, we present a highly efficient and secure GKM protocol for dynamic IoT settings, which maintains forward and backward secrecy at all times. Our proposed protocol uses only symmetric encryption, and is  completely resistant to  collusion attacks. Also, our protocol is highly flexible and can handle several new scenarios in which device or user dynamics may take place, e.g., allowing a device group
to join or leave the network or creation or dissolution of a
user group, which are not handled by schemes proposed in prior literature. We evaluate the performance of the proposed protocol via extensive mathematical analysis and numerical computations, and show that it outperforms the
GroupIt scheme in terms of the communication and computation costs incurred by users and devices.  
\end{abstract}


\section{Introduction}
The number of Internet connected entities was approximately 27 billion in 2019, and is expected to rapidly reach the enormous number of 75 billion by 2025~\cite{Maayan}. The modern Internet is a global network of intelligent devices ranging from traditionally connected devices such as desktop and laptop computers and smartphones to objects of daily use such as watches and spectacles, electronic home appliances (even window shutters), sensing and actuating devices such as patients' medical sensors and industrially deployed temperature sensors and machinery~\cite{Das1}. The introduction of IPv6 addresses~\cite{Jiang} has made it possible for all these devices to connect to the Internet and has helped in automation of tasks, decreased the need for human intervention, increased accessibility and convenience in functionality and data acquisition. The ability to connect resource-constrained devices such as sensors and actuators as well as everyday objects to the Internet has led to the notion of the \emph{Internet of Things} (IoT)~\cite{Weber}.

Compared to traditional devices such as desktop computers and laptops, IoT devices have low computational and storage power. As a result, they are not able to run the protocols traditionally used to achieve secure communications in the Internet, e.g., those that use public key cryptography~\cite{Granjal}. At the same time, confidentiality and message-integrity of data communicated by and to IoT devices is essential. It may relate to the geographical coordinates of a user, medication provided to a patient, customized use of home appliances, parameters and commands exchanged by industrial machinery or security alarms. This broad range of essential functions, without sufficient security, can potentially result in huge losses. Hence, it is crucial to design effective security mechanisms for resource-constrained IoT devices. A review of prior research literature in which security mechanisms for the IoT are proposed is provided in Section~\ref{RW}. 

Many present-day scenarios require communication to and data acquisition from multiple devices with similar functionalities, e.g., medical sensors, Industrial Wireless Sensor Network nodes and nodes installed in smart homes. For such scenarios, \emph{group communication} in the form of multicasting and broadcasting has proven to be effective~\cite{Veltri}. \emph{Group Key Management} (GKM) involves the handling, revocation, updation and distribution of cryptographic keys to members of various groups in a communication network~\cite{Rafaeli}. GKM can be adopted in a centralized or distributed manner~\cite{Omar}. Centralized GKM schemes involve a centralized entity like a Key Distribution Center, which generates keys and performs the aforementioned functions. On the other hand, in distributed GKM schemes, nodes are clustered into several groups and cluster heads are assigned to manage keys locally~\cite{Esposito}. However, the latter approach incurs high computational costs; as a result, most research, including this paper, is based on the centralized approach. 

Now, classical GKM schemes perform inefficiently in \emph{dynamic} IoT environments, which are those wherein nodes frequently leave or join a network or migrate from one group to another over time~\cite{Kung}. In such scenarios, whenever changes occur in the network, keys need to be updated and distributed in such a way that \emph{backward and forward secrecy}~\cite{Rafaeli} is maintained. Apart from this, the use of asymmetric cryptography in dynamic IoT environments limits the overall scalability of the system. Therefore, a GKM scheme based on symmetric cryptography is needed that can scale well in dynamic settings. 

One major goal of such a GKM scheme is to exercise access control over the network~\cite{Kung}. Specifically, a set of users, also called subscribers, send data to and receive data from an IoT device. For example, these users may be nursing staff appointed for a patient, who wish to obtain periodic vital sign data from medical sensors deployed on the patient's body. It is of utmost importance to provide confidentiality and integrity of messages, as well as avoid any unauthorized access. Sharing a group key between a device and its subscribers is an effective way to impose such access control. However, as  users join, leave or migrate between groups (e.g., a nurse is allocated to or deallocated from a patient or allocated to a different patient), real-time GKM is desired to fulfill the security requirements.

In this paper, we present a centralized GKM scheme for dynamic IoT environments, which maintains forward and backward secrecy at all times. Our scheme only employs resource-constrained network friendly symmetric encryption, due to which the centralized controller can efficiently handle a large number of subscription changes, node entries and exits at any given time. Moreover, our scheme is designed in such a way that no user outside the network can collude with a registered device and deduce group secrets. In our scheme, we have also made provisions for new scenarios in which device or user dynamics may take place, e.g., allowing a device group to join or leave the network or creation or dissolution of a user group, which were not provided by schemes proposed in prior literature~\cite{Kung}; this makes our proposed scheme more flexible in handling dynamics. Finally, via mathematical analysis and numerical computations, we evaluate the performance of the proposed scheme and show that it outperforms the GKM scheme proposed for dynamic IoT environments in prior work~\cite{Kung}.

The rest of the paper is organized as follows. In Section~\ref{RW}, we review the prior research literature on security in the IoT and GKM. Section~\ref{model} describes our system model, problem formulation and reviews some relevant background. In Section~\ref{GroupIt}, we briefly describe the GKM scheme `GroupIt' proposed for dynamic IoT environments in~\cite{Kung} and identify its limitations. Our proposed GKM scheme is described in Section~\ref{PP}. In Section~\ref{performance}, a mathematical analysis of the performance of the proposed scheme in terms of storage, computation and communication costs is provided. In Section~\ref{numerical}, we provide a numerical computations based performance evaluation of the proposed scheme. Finally, we provide conclusions and directions for future research in Section~\ref{conclusion}.


\section{Related Work}
\label{RW}
Security in the IoT has gained significant attention in the research community--  surveys of various enabling technologies for the IoT (e.g., RFID, Zigbee, MQTT, CoAP), including security aspects, are provided in~\cite{Granjal},~\cite{Lin},~\cite{Sicari}.  The IEEE standard 802.15.4 defines rules for the physical and medium access control (PHY and MAC) layers, aimed at providing low cost, low power and low data rate wireless connectivity among constrained devices~\cite{Gutierrez}. The Internet Engineering Task Force (IETF) has also developed protocols for various layers to support the IoT~\cite{Sheng}.   IPv6 over Low power Wireless Personal Area Networks (6LoWPAN)~\cite{Schrickte}, an IETF standard, is one of the technologies adopting 802.15.4. It acts as a gateway between the Internet and constrained devices, performing various tasks such as header compression, fragmentation and reassembly of IPv6 packets from the Internet into sizes that can be sent over constrained networks using 802.15.4. The Routing Protocol for Low Power and Lossy Networks (RPL) is another IETF standard, designed for routing over constrained networks. DTLS header compression and TLS-DTLS mapping~\cite{Yang} have also been proposed to increase the allowed application-layer payload.
The design of secure protocols for various IoT standards is a topic of recent research~\cite{Granjal}.  

Multifactor authentication has been extensively studied in the past decade and a plethora of schemes have been proposed for various IoT settings. In~\cite{Turkanovic}, a novel $2$-factor authentication scheme for any remote user to securely connect with a given sensor node is presented. Wireless Sensor Networks (WSNs) form an integral part of the IoT. In~\cite{Jiang}, the authors propose an efficient and lightweight $3$-factor authentication and key establishment scheme, based on Biohashing and the Rabin Cryptosystem, for WSNs to securely connect to the Internet. Elliptic Curve Cryptography has also found novel applications in designing keying protocols for constrained network architectures~\cite{Porambage,Das,Chang,Ferrari}. Online validation of public key certificates is another problem in the field of IoT security. In~\cite{Park}, the authors introduced a protocol that employs a cryptographically generated address for bootstrapping to secure the join and certificate issuance mechanisms.

Over the years, several highly efficient \emph{Group Key Establishment} (GKE) schemes have been proposed. GKE refers to one-time authentication and key agreement among two or more parties in a group who wish to exchange messages. In~\cite{Eldefrawy}, a lightweight key distribution protocol for Industrial IoT, based on the Chinese Remainder Theorem, which requires only a single message transmission, and provides backward and forward secrecy is proposed. In~\cite{Robinson}, the authors explored the use of Physical Unclonable Functions with appropriate guarantees as a tool for GKE. In~\cite{Ferrari1}, a context based GKE protocol, in which devices calculate a fingerprint from their surrounding context and generate a shared secret among them, was proposed. 

However, recent literature has shown a shift towards Group Key Management (GKM) in the IoT. Complementing GKE, GKM deals with the updation, revocation and distribution of established cryptographic keys in a group. In~\cite{Rafaeli}, a broad survey and comparison of several centralized, decentralized and distributed GKM protocols was provided. The authors compared various mechanisms proposed in the research literature for GKM, including LKH, OFT, ELK and MARKS. In~\cite{Veltri}, the authors proposed a novel batch based centralized approach to efficiently manage group keys in generic ad hoc networks, while limiting the computational and communication costs due to group membership changes caused by users  joining and leaving. They handle such membership changes and events of explicit membership revocation at different pre-assigned times by partitioning time into fixed length intervals. Our scheme, on the other hand, allows more flexibility for users by removing the need to alter memberships at pre-assigned times. In~\cite{Omar}, a Logical Neighbor Tree (LNT) based approach to distribute key update messages in the events of membership changes is proposed. The proposed scheme reduces the key update latency from $O(n)$ to $O(\log(n))$, where $n$ is the number of group members. Unlike equally ranked group entities in our protocol, their scheme includes a group controller for each group, which is responsible for local security inside the respective groups. 

In~\cite{Wong}, the Logical Key Hierarchy (LKH) approach in which keys are arranged in a hierarchy and a Key Distribution Center maintains and updates all the keys was proposed. It makes use of symmetric key encryption to multicast key update messages to group members, which are located at the leaf nodes of a logical tree; the communication cost of this scheme is of the order of $O(\log(n))$, where $n$ is the number of nodes in the network. As we will see later, we have also used LKH as a basic mechanism to form groups in our model. Later,~\cite{Son} presented a Topological Key Hierarchy (TKH) scheme for WSNs as an alternative to LKH. TKH generates a key tree by using the underlying sensor network topology and considering subtree-based key tree separation. A basic difference from LKH is that topologically adjacent entities share similar key material, and therefore, receive similar rekeying messages. 
Recently in~\cite{Mughal}, the authors proposed a logical tree based secure mobility management scheme (LT-SMM) using mobile service computing for the IoT. The model includes a group head for each group, which is computationally more capable than other nodes in the group. Although using the LKH scheme as in our proposed protocol, their main focus is on reducing excessive rekeying when a node only migrates from one group to another--  coordinating the group heads through the Base Station helps in exchanging key material, and completing the migration process with a small overhead for the migrating node.

The closest to our research is the work~\cite{Kung}, which, to the best of our knowledge, is the only paper in prior work that proposes a GKM scheme for dynamic IoT environments. In~\cite{Kung}, a system model that consists of several groups of users, which selectively subscribe to a desired subset of device groups and collect data, was considered. The authors proposed a two-tier GKM scheme, GroupIt, to handle frequent subscription changes. They leveraged the LKH scheme and the Chinese Remainder Theorem (CRT) to reduce the communication overhead incurred due to node entry and exit events. While the CRT manages keys shared between the centralized controller and groups, LKH handles key distribution within groups. In~\cite{Kung}, it is claimed that the GroupIt scheme  achieves forward and backward secrecy, as well as a low probability of collusion attacks. However, in this paper, we propose a scheme that outperforms the GroupIt scheme in  several respects; in particular, our scheme completely eliminates collusion attacks and is more efficient since, in contrast to the GroupIt scheme, it does not use asymmetric encryption. Moreover, our proposed GKM scheme is  more flexible than GroupIt since the former handles new scenarios in which device or
user dynamics may take place, e.g., allowing a device group to join or leave the network or creation or dissolution of a user group; GroupIt does not explicitly handle any of the above scenarios. Finally, we show via a combination of mathematical analysis and numerical computations that our proposed scheme outperforms GroupIt in terms of the computation and communication cost incurred by users as well as devices.

\section{Model, Problem Formulation and Background}
\label{model}
We describe the system model and adversarial model in Sections~\ref{model:A} and~\ref{model:B}, respectively. Then in Section~\ref{model:C}, we provide a brief overview of the LKH scheme, which we use as a part of our proposed scheme.

\subsection{System Model}
\label{model:A}
The system consists of a Key Distribution Center (KDC), IoT devices and subscribers (users). Henceforth, the terms ``user'' and ``subscriber'' are used interchangeably. The KDC is a trusted centralized entity that generates, distributes and updates key material for all devices and users. The KDC is assumed to have high computational and storage capability; hence, while quantifying the protocol's efficiency, we focus on the computational and storage costs incurred by the devices and users. Each IoT device collects the required data and periodically sends it to its subscribers. For instance, devices (sensors) attached to a patient's body may periodically send the patient's vital signs to their subscribers (authorized nurses).

One or more devices of similar functionality or security classification are grouped together into a ``\emph{Device Group}'' (DG). Also, one or more users are grouped together into a ``\emph{Subscriber Group}'' (SG).  Each user (respectively, device) belongs to exactly one SG (respectively, DG).
Each SG is subscribed to one or more DGs. In particular, a SG is  a group of users subscribing to a specific subset of DGs. Thus, if there are $P$ DGs, then there are $2^{P}-1$ possible SGs, corresponding to different non-empty subsets of the collection of the $P$ DGs. 

\begin{figure}
  \centering
  \includegraphics[scale=0.32]{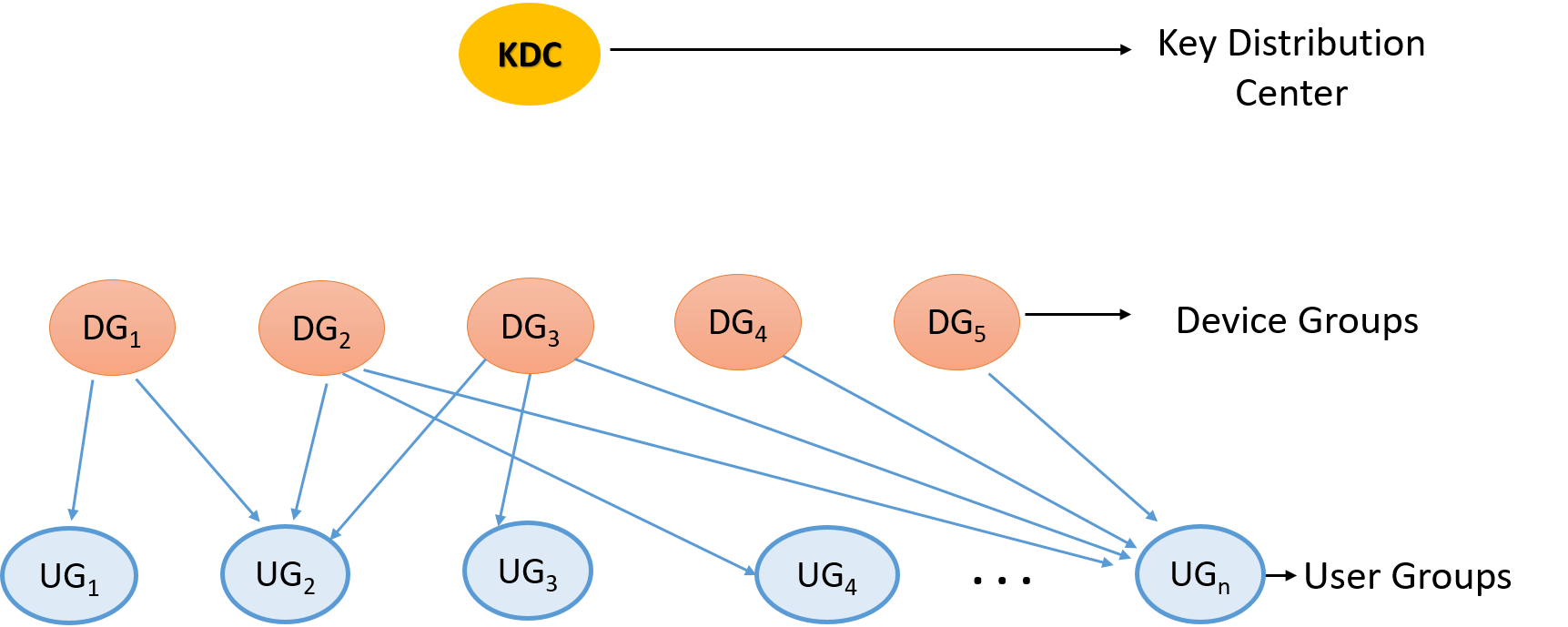}
  \caption{There are five DGs, DG\textsubscript{1}, \ldots, DG\textsubscript{5}, and $n$ UGs, UG\textsubscript{1}, \ldots, UG\textsubscript{n}.  UG\textsubscript{1} subscribes only to DG\textsubscript{1}, UG\textsubscript{2} subscribes to DG\textsubscript{1}, DG\textsubscript{2} and DG\textsubscript{3}, and so on.}
  \label{fig1}
\end{figure}

Each user or device in the network has to authenticate itself and have a secret key established with the KDC before becoming part of any group. Also, each user or device group has its own group key, initially provided by the KDC. Each device can exchange encrypted data with its subscribed users using a key that is known only to the KDC and the communicating parties, called the ``\emph{device key}'' of the device. Fig.~\ref{fig1} illustrates the system model. 

Our objective is to design a GKM scheme that enables efficient updation or revocation of the cryptographic keys assigned to different users and devices in the network when changes take place in the network such that \emph{forward and backward secrecy} is maintained at all times. Recall that by forward (respectively, backward) secrecy, we mean that a leaving (respectively, joining) user is prevented from decrypting messages exchanged after (respectively, before) it leaves (respectively, joins) the group~\cite{Rafaeli}.  The possible scenarios in which key material needs to be updated or revoked are as follows: A user or a device joining or leaving the network, addition or removal of a DG, a user occupying an empty SG and an existing user leaving the last spot in a SG. Also, we seek to design a GKM scheme that minimizes the computational, communication and storage costs incurred to the users and devices in the network when changes take place in the network. 

Similar to most prior works in GKM, we make the following assumptions. First, all network entities use the same cipher suite and keys are sufficiently large~\cite{Porambage}. Second, the system is reactive to tampering; therefore, any node capture or compromise will be detected in practically small time, resulting in appropriate revocation and updation of device and user key material. Different approaches such as mobility based~\cite{Conti} and control theoretic~\cite{Bonaci} modelling have been presented in prior work to detect and revoke compromised nodes. Finally, we assume that the message integrity of all communication exchanged during the network operation is protected using standard mechanisms (e.g., message authentication codes)~\cite{Granjal}.

\subsection{Adversarial Model}
\label{model:B}
In our model, the adversary may be a node outside the network, a malicious user belonging to one of the user groups, or a corrupted device that tries to access data belonging to other devices. If the adversary is a user belonging to the network, it may try to either compromise forward or backward secrecy or access the data trasmitted by devices to which it is not subscribed. 
The adversary aims to breach access control and decrypt the periodic data transmitted by devices, which it is not authorized to access, either by illegitimately gaining access to key material or colluding with devices existing in the network.

\subsection{Logical Key Hierarchy (LKH)}
\label{model:C}
LKH is a centralized GKM technique designed for achieving secure and efficient rekeying and message transmissions within a group~\cite{Wong}, while providing forward and backward secrecy. LKH provides methods to update the cryptographic keys shared within a group when a node joins or leaves the group. Moreover, it helps to multicast messages to different nodes in the group while incurring a low communication cost.

\begin{figure}
  \centering
  \includegraphics[scale=0.28]{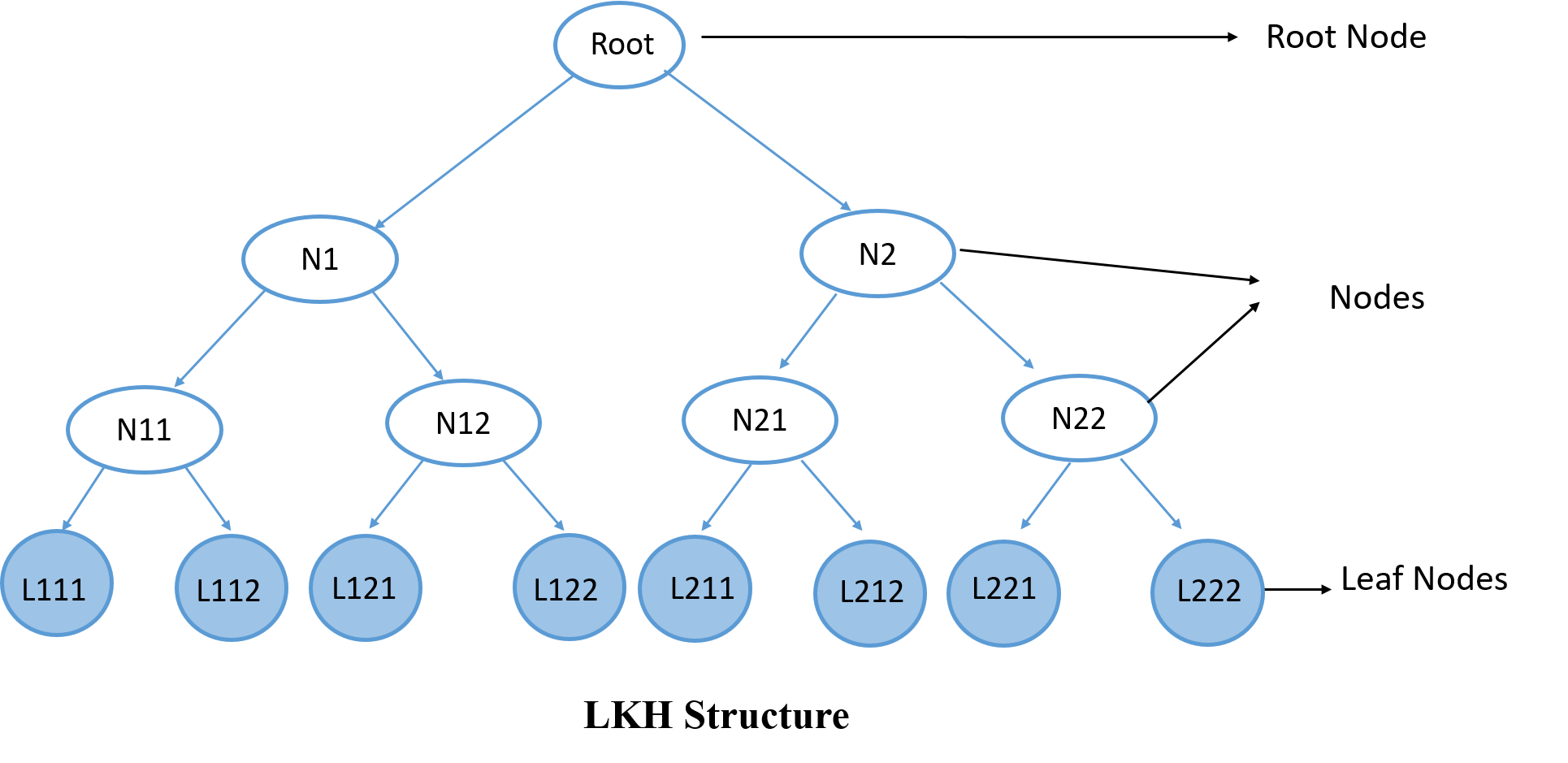}
  \caption{The figure shows an example LKH tree structure.}
  \label{fig2}
\end{figure}

Under the LKH scheme, a KDC maintains a tree of keys~\cite{Rafaeli} as shown in Fig.~\ref{fig2}. A group arranged in an LKH tree structure has all its members occupying the leaf nodes of the tree. Each node of the tree other than the leaves and the root is associated with a key called the \emph{Key Encryption Key} (KEK). Also, the root node is associated with a key called the \emph{group key}. A member knows all the KEKs belonging to its ancestors in the tree and the group key. For example, node L122 in Fig.~\ref{fig2} knows the KEKs corresponding to N12 and N1 and the group key. Note that each member stores at most $\lceil \log_2(k) \rceil$ KEKs, where $k$ is the number of group members.
Apart from this, every member also has a secret key shared with the KDC. 

An LKH tree structure helps to reduce the communication cost incurred in sending multicast messages from the KDC to group members. In the example in Fig.~\ref{fig2},  if a message is to be sent to the nodes L111, L112, L121 and L122, then a single multicast encrypted with the KEK corresponding to the common ancestor N1 can be sent under LKH; this is significantly more efficient than the transmission of four separate unicast messages encrypted using the different secret keys shared between the KDC and the four recipient nodes.  

\begin{figure}
  \centering
  \includegraphics[scale=0.28]{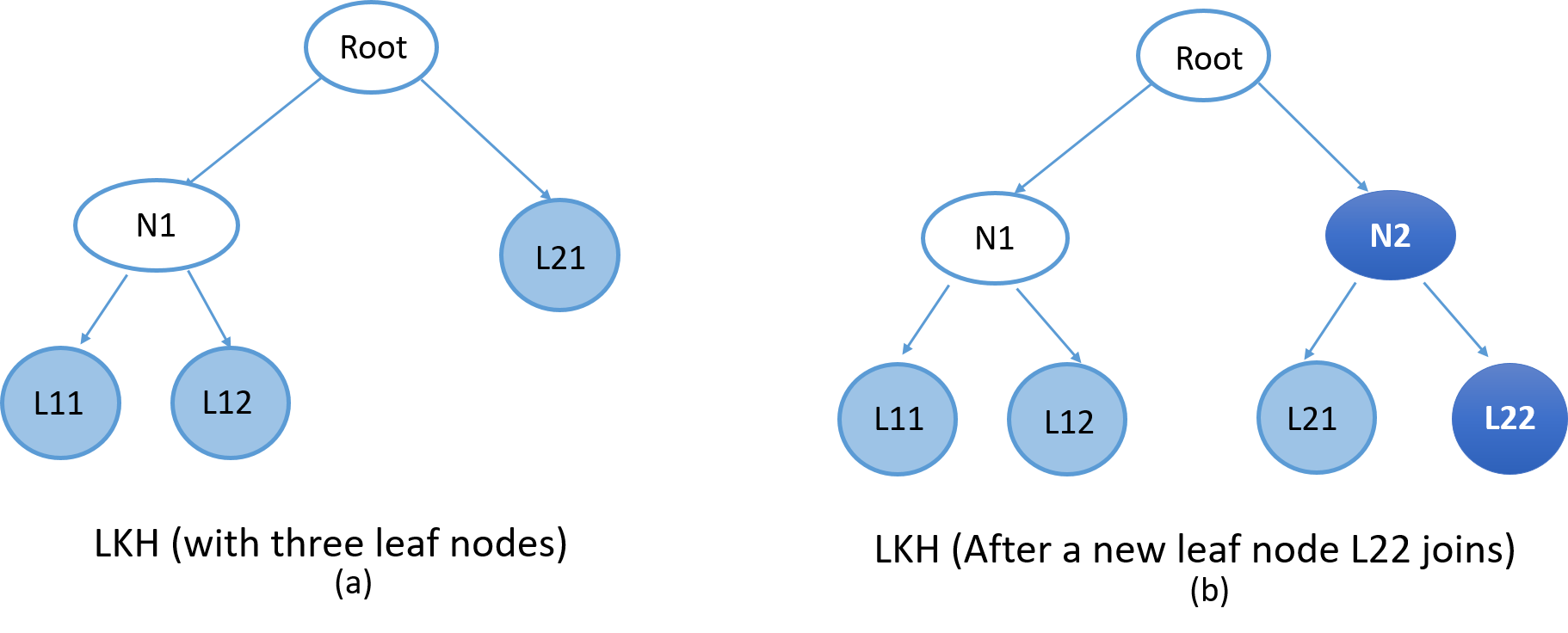}
  \caption{The figure shows the LKH tree before and after node L22 joins the group.}
  \label{fig3}
\end{figure}

\begin{figure}
  \centering
  \includegraphics[scale=0.24]{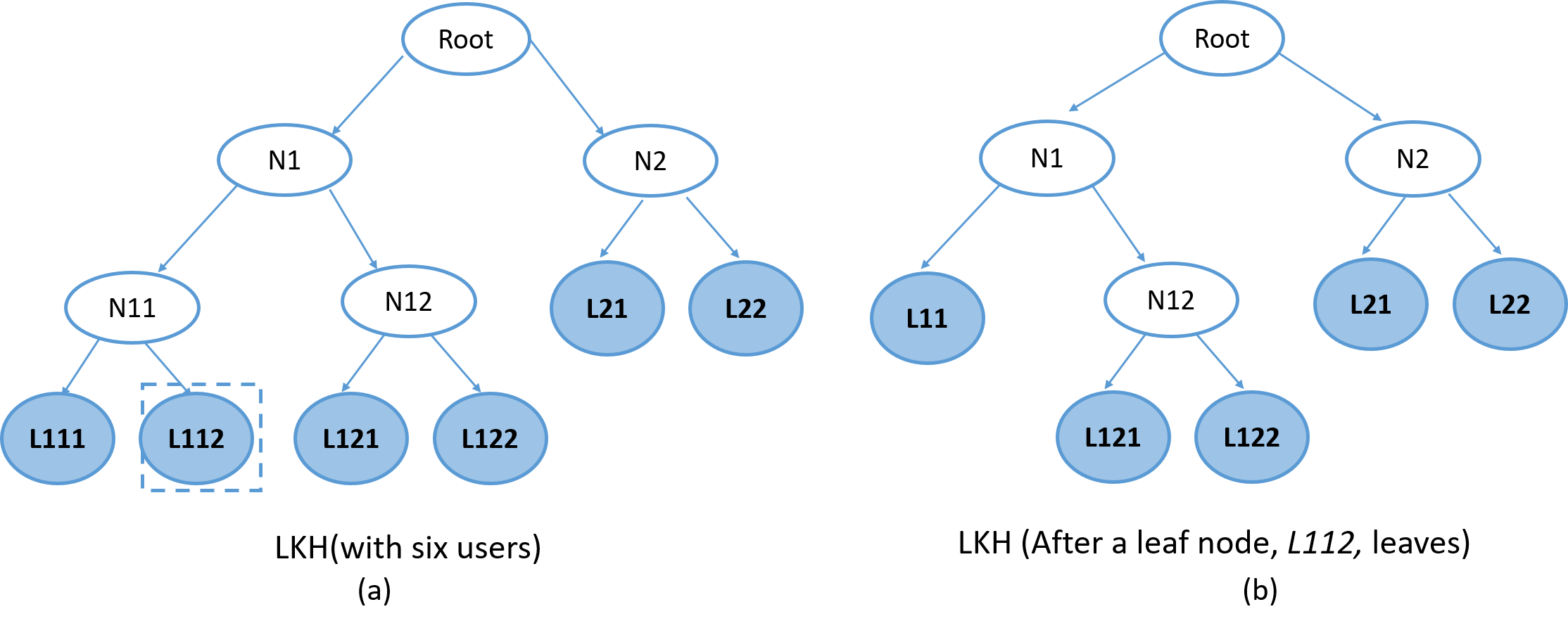}
  \caption{The figure shows the LKH tree before and after node L112 leaves the group.}
  \label{fig4}
\end{figure}

The KDC is responsible for updating the group secrets in the events of member entry or exit so as to maintain backward and forward secrecy. Consider the LKH tree shown in Fig.~\ref{fig3}(a). When a new node L22 joins the group, the KDC first establishes a secret key with it. This is followed by creation or updation  of all the keys corresponding to the ancestors of the joining node L22 in the new LKH tree shown in Fig.~\ref{fig3}(b)  (i.e., the KEK corresponding to N2 and the group key) so as to maintain backward secrecy. The KDC then sends the newly created KEK of N2 and the updated group key  to L22 using the secret key shared between the KDC and L22, and to other affected members who share a common ancestor with L22 using efficient multicasts using shared KEKs or secret keys. In this example, the KDC sends the new KEK of N2 and the group key to L21 using the shared secret key between the KDC and L21, and sends the updated group key to L11 and L12 using the KEK corresponding to N1. 

Similarly, if an existing member leaves the group, as shown in Fig.~\ref{fig4}, all its ancestral KEKs and the group key need to be updated to maintain forward secrecy. The KDC sends the updated KEKs and/ or the group key to the remaining members using suitable shared KEKs or secret keys. In this example, after L112 leaves, the KDC sends the updated KEK corresponding to N1 and the group key to L11 using the shared secret key between the KDC and L11, and to L121 and L122 using the KEK corresponding to N12. It also sends the updated group key to L21 and L22 using the KEK corresponding to N2.


\section{Review and Limitations of GroupIt Scheme}
\label{GroupIt}
In Section~\ref{GroupIt:A}, we briefly review the GroupIt scheme proposed for dynamic IoT environments in~\cite{Kung} and in Section~\ref{GroupIt:B}, we discuss its limitations.

\subsection{GroupIt}
\label{GroupIt:A}
GroupIt is a scheme designed for GKM in a dynamic IoT environment. The system model for which GroupIt is designed is the same as the one described in Section~\ref{model:A}. GroupIt uses LKH and the Chinese Remainder Theorem (CRT) to limit the communication overhead incurred for key updation due to node entry and exit events. While the CRT manages keys between the KDC and groups, LKH handles key distribution within a group. 

To initialize, the KDC generates secret keys for all users and devices and the LKH KEKs and group keys for all SGs and DGs, and distributes them to the respective users and devices. Each device is associated with a unique ID, and each DG is associated with a secret Traffic Encryption Key (TEK), both of which are known to each user subscribed to the device. In this way, the user can calculate the device key, which is a function of the device ID and TEK of the device group, for each device to which it is subscribed.  The authors enlist four scenarios, discussed below, in which key material (group key, KEKs, TEKs) needs to be updated, and provide the corresponding steps to achieve forward and backward secrecy. For all these cases, LKH is used to update the group key and KEKs within a SG or DG.

\subsubsection{New User (U\textsubscript{i}) Joins a Group SG\textsubscript{x}}
First, the KDC establishes a secret key with the new user. It also notifies all the DGs subscribed by SG\textsubscript{x}, and the users subscribed to those DGs to hash-update their TEKs. Hash update refers to the application of a cryptographic hash function~\cite{Granjal} to the TEK, so that backward secrecy can be achieved. Then, the KEKs within SG\textsubscript{x} are updated as per the LKH join algorithm described in Section~\ref{model:C}. Finally, the new user is provided the required key material using the secret key established between the KDC and the new user. 

\subsubsection{Existing User (U\textsubscript{i}) Leaves a Group SG\textsubscript{x}}
The KDC first updates the group key of SG\textsubscript{x}. Then, for each DG that was subscribed to by the leaving user, it broadcasts to the other users, updated TEKs and methods to update the device IDs of the devices, using double encryption, i.e., two layers of encryption, first using an asymmetric key on the message and then encrypting the output with a symmetric key TEK. Finally, the new TEKs and device IDs are sent to the devices subscribed to by the leaving user.

\subsubsection{New Device Joins a Group}
First, the KDC establishes a secret key with the joining device. Then, it updates the group key and the KEKs of its DG. Finally, the ID of the new device is shared with the users subscribed to it.

\subsubsection{Existing Device Leaves a Group}
The KDC first broadcasts over the network that the device is no longer available, and then updates the group key and KEKs for the other devices in the group.

\subsection{Limitations of GroupIt}
\label{GroupIt:B}
\subsubsection{Employment of Asymmetric Encryption}
When an existing user leaves a SG, the KDC broadcasts doubly encrypted messages to the other users, each containing update methods for a DG subscribed to by the leaving user. The inner layer of this encryption requires asymmetric decryption by the other users. This has a significant computational cost~\cite{Wang}.     

\subsubsection{Collusion Attacks}
\label{SSSC:GroupIt:B:collusion}
The device keys are derived from device IDs and TEKs of the DGs. If the device ID is not updated when a user subscribed to the device leaves, and the leaving user colludes with a device having the new TEK, then they can combine their knowledge to derive the other devices' new device keys. GroupIt has a non-zero probability that these device IDs will not be updated when a user leaves. 

\subsubsection{Erroneous Steps of Key Updation}
To achieve backward secrecy, the LKH group key of the SG must be updated when a new user joins the SG. But the scheme does not account for such a key update. Also, when an existing user leaves, the KEKs of the SG are not updated. Due to this, the new group key can be compromised by  being revealed to the leaving user. Moreover, in the double encryption performed when a user leaves, the old TEK is available with the leaving user and it does not add to the forward secrecy of the system, since the user can easily decrypt the outer layer of encryption even after leaving.

\subsubsection{Low Flexibility}
GroupIt does not explicitly handle the cases when a new DG joins, an old DG leaves, a new user occupies an empty SG, and an existing user leaves the last occupied spot in a SG.



\section{Proposed GKM Scheme}
\label{PP}
Our proposed GKM scheme is described in the following subsections.

\subsection{Setup and Pre-deployment}
\label{PP:1}
\begin{figure}
  \centering
  \includegraphics[scale=0.26]{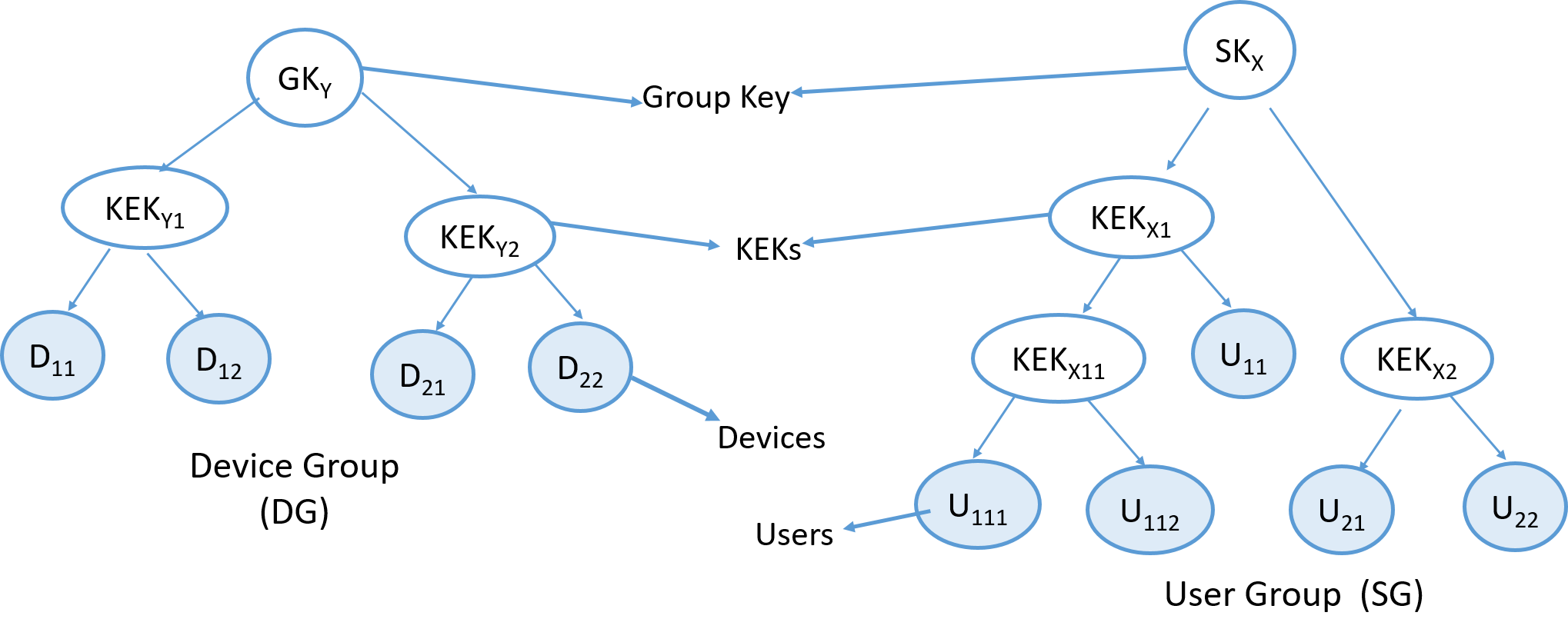}
  \caption{The LKH tree structure of four devices in a DG (left) and five users in a SG (right).}
  \label{fig5}
\end{figure}

\begin{figure}
  \centering
  \includegraphics[scale=0.26]{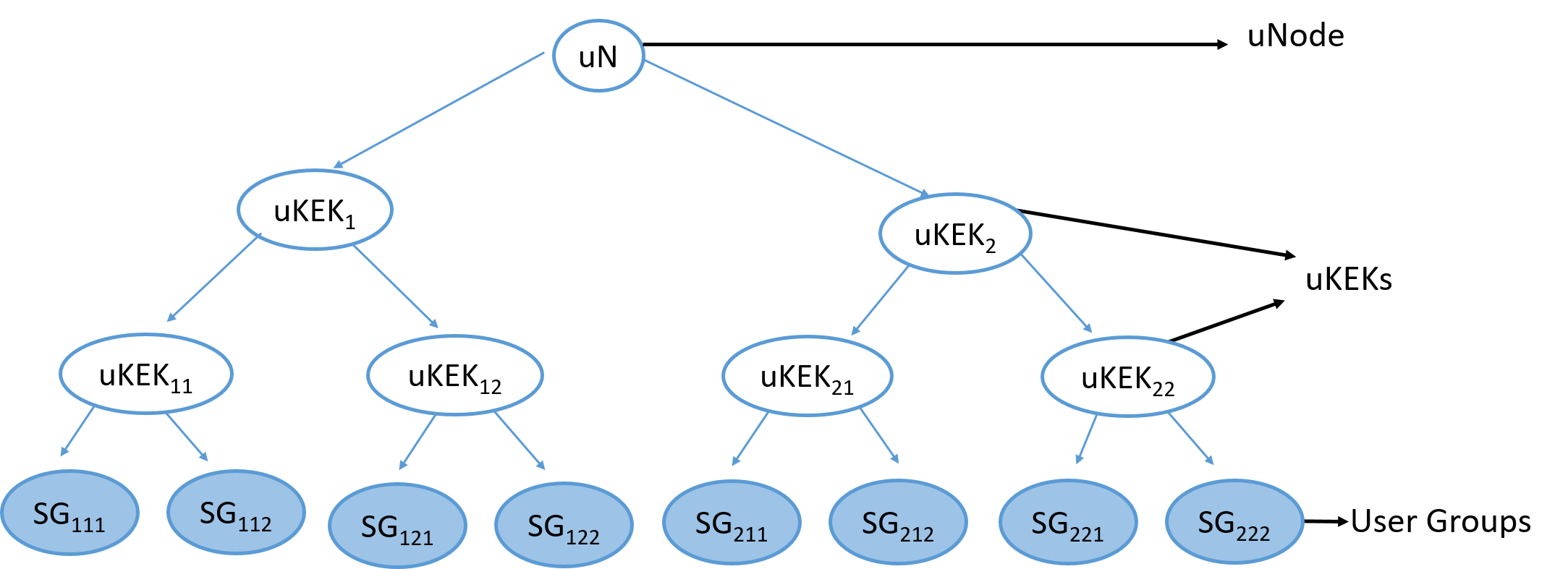}
  \caption{The outer LKH tree structure of 8 SGs.}
  \label{fig6}
\end{figure}

Users (respectively, devices) in a SG (respectively, DG) are arranged in an LKH tree structure and occupy the leaf nodes of the tree (see Fig.~\ref{fig5}). Also, the SGs themselves are part of a bigger (outer) LKH structure, which we refer to as the `uLKH', as shown in Fig.~\ref{fig6}; we refer to a KEK in the outer LKH structure as a `uKEK'. When we say that a SG is assigned a set of uKEKs, it means that each user in the SG is assigned that set of uKEKs. In the example in Fig.~\ref{fig6}, all users in SG\textsubscript{111} and SG\textsubscript{112} have uKEK\textsubscript{11}, uKEK\textsubscript{1} and the group key, uN, corresponding to the root, uNode. Note that uN can be used to securely broadcast a message to all the users in the network. The group key,  uN, and the uKEKs together enable the KDC to multicast messages to multiple SGs. In the example in Fig.~\ref{fig6}, if a message is to be sent to SG\textsubscript{111}, SG\textsubscript{112}, SG\textsubscript{121} and SG\textsubscript{122}, then the KDC will simply encrypt it with uKEK\textsubscript{1} and multicast the message to the four groups. 

The device key for a device $j$ is defined as:
\begin{equation}
\label{EQ:E:m:j}
DK_j=h(ID_j||n_j)
\end{equation}
where $h(\cdot)$ is a cryptographic hash function, $ID_j$ is a secret  and $n_j$ is a nonce, both $ID_j$ and $n_j$ are shared between device $j$ and the KDC, and $||$ denotes a concatenation. 

Each device $j$ encrypts messages containing its periodic data using $DK_j$ and multicasts it to all the SGs subscribed to device $j$. Each user subscribed to a device $j$ knows its device key $DK_j$.

At the beginning of the network operation, the KDC performs the following actions:
\begin{itemize}
  \item Establishes secret keys with each user and device. 
  \item Generates and distributes the group key GK\textsubscript{y} for each device group DG\textsubscript{y}.
  \item Generates and distributes the group key SK\textsubscript{x} for each subscriber group SG\textsubscript{x}.
  \item Generates and distributes the KEKs for the LKH tree of each SG and DG.
  \item Generates the device IDs and nonces for all devices, distributes them to the respective devices and computes the device keys for all devices.
  \item Distributes the device key of each device $j$ to all the SGs subscribed to device $j$.
  \item Generates and distributes the uKEKs and the group key for the outer SG tree. 
\end{itemize}

\subsection{When a New User Joins}
\label{PP:2}
When a new user $U_i$ joins a non-empty user group SG\textsubscript{x}, the following actions are performed:  
\begin{itemize}
  \item The KDC establishes a secret key with the new user $U_i$. 
  \item The KDC sends a broadcast message that all devices from the DGs  to which SG\textsubscript{x} subscribes and the old users from SGs subscribed to those DGs should hash update the device keys, say $DK_j$, and that the users in SG\textsubscript{x} should hash update the group key SK\textsubscript{x}. The above updates are performed as follows:
  \begin{enumerate}
      \item $DK^{\prime}_j = h(DK_j)$
      \item $SK^{\prime}_x = h(SK_x)$
  \end{enumerate}
This ensures backward secrecy.
  \item The KDC sends the updated KEKs, say $KEK^{\prime}$s,  corresponding to the ancestors of $U_i$ in the new LKH tree of SG\textsubscript{x} to the members of SG\textsubscript{x} encrypted by the appropriate secret keys or shared KEKs. 
  \item The KDC sends the updated uKEKs, say $uKEK'$s, of the outer LKH tree  to the SGs having a common ancestor with SG\textsubscript{x}, encrypted by the appropriate group keys or shared uKEKs. 
  \item The KDC shares the relevant secret information ($DK^{\prime}_j$, $SK^{\prime}_x$, $KEK^{\prime}$s, $uKEK'$s) with the new user $U_i$. 
\end{itemize}

\subsection{When an Existing User Leaves}
\label{PP:3}
Suppose an existing user leaves its SG, say SG\textsubscript{x}, but SG\textsubscript{x} continues to be non-empty. Then a simple hash updation of secret information will not suffice, and therefore, the following procedure is adopted. The exit of the user not only affects the SG, SG\textsubscript{x}, from which the user left, but also the SGs that subscribe to the devices to which the leaving user subscribed, called the ``concerned SGs''. As a result, we first update the uKEKs in the outer LKH tree and then send the updated device keys using them to the concerned SGs.  The following actions are performed:
\begin{itemize}
  \item The KDC sends a message to the devices subscribed to by the leaving user to update their nonces. The nonce can be simply incremented by one, i.e., $n_j^{\prime} = n_j + 1$. This update does not require any unicast communication between the KDC and the devices, thus helping in limiting the communication cost. The new device key is calculated in the same way as before by the devices: \[DK'_{j} = h(ID_{j}||n'_{j})\]  
This helps in achieving forward secrecy. 
  \item The KDC sends the updated KEKs for the users having  a common ancestor with the leaving user in the LKH tree of SG\textsubscript{x}  and the updated group key, $SK_x^{\prime}$,   to the appropriate members of SG\textsubscript{x}, encrypted by appropriate secret keys or KEKs. 
  \item The KDC sends the updated uKEKs to the SGs having a common ancestor with SG\textsubscript{x} in the outer LKH tree, encrypted by the appropriate group keys or shared uKEKs. 
  \item For each DG subscribed to by SG\textsubscript{x}, the KDC sends a multicast containing the updated device keys to the SGs subscribed to the DG, using the appropriate updated uKEKs. 
\end{itemize}

\subsection{When a New Device Joins}
\label{PP:4}
The new device, say $D_k$, is added to one of the DGs, say DG\textsubscript{y}. The KDC creates a secret identity, $ID_k$, and a secret nonce,  $n_k$, for $D_k$. The  following actions are performed:
\begin{itemize}
  \item The KDC establishes a secret key with the device, provides it with $ID_k$ and $n_k$, and both the KDC and the device $D_k$ derive the device key as follows: \[DK_{k} = h(ID_{k}||n_{k}).\]
  \item The KDC sends a message to the existing devices in DG\textsubscript{y} to hash update their group key: $GK'_y = h(GK_y)$. This helps in achieving backward secrecy. 
  \item The KDC sends the updated KEKs corresponding to the ancestors of $D_k$ in the LKH tree of DG\textsubscript{y} to the appropriate devices in DG\textsubscript{y}, encrypted by appropriate secret keys or shared KEKs. 
  \item The KDC securely multicasts the device key $DK_k$ to the SGs subscribed to DG\textsubscript{y} using the appropriate uKEKs.  
\end{itemize}

\subsection{When an Existing Device Leaves}
\label{PP:5}
Suppose an existing device, say $D_k$, leaves a DG, say DG\textsubscript{y}, but  DG\textsubscript{y} continues to be non-empty. Then the following actions are performed:
\begin{itemize}
  \item The KDC sends a broadcast message to all SGs indicating that device $D_k$ is no longer available. This broadcast message is sent using the group key of the outer LKH tree, i.e., the key, uN, corresponding to the root uNode.   
  \item The KDC sends the updated KEKs and the updated group key,  $GK_y^{\prime}$, of the LKH tree of DG\textsubscript{y} to the appropriate members of DG\textsubscript{y}, encrypted using the appropriate secret keys or shared KEKs. 
\end{itemize}

\subsection{When a New User Occupies an Empty Group}
\label{PP:6}
When a new user, say $U_i$, joins an empty SG, say SG\textsubscript{x}, the following actions are taken:
\begin{itemize}
  \item The KDC establishes a secret key with the user $U_i$.  
  \item The KDC creates a new LKH tree for the group SG\textsubscript{x} and
derives its group key SK\textsubscript{x}. 
  \item The KDC sends a broadcast message announcing that all devices to which SG\textsubscript{x} subscribes and the old users subscribed to those devices should hash update the device keys: $DK'_j = h(DK_j)$. This helps in achieving backward secrecy. 
  \item The KDC modifies the outer LKH structure, uLKH, to add SG\textsubscript{x} to it; in particular, the KDC sends the updated uKEKs, say $uKEK'$s, and the updated group key, say $uN'$, corresponding to uNode to the SGs having a common ancestor with SG\textsubscript{x}, encrypted by the appropriate group keys or shared uKEKs. 
  \item The KDC shares the relevant secret information ($DK_j'$, SK\textsubscript{x},  $uKEK'$, $uN'$) with the user $U_i$, by encrypting it using the secret key shared between the KDC and user $U_i$.  
\end{itemize}

\subsection{When an Existing User Leaves the Last Spot in a SG}
\label{PP:7}
When an existing user leaves the last spot in a SG, say SG\textsubscript{x}, the LKH tree keys of the SG need not be updated since new keys are derived if and when the SG starts re-populating. Nevertheless, we need to update the uKEKs in the outer LKH tree and then send device keys using them to the concerned SGs as in Section~\ref{PP:3}. The following actions are performed:
\begin{itemize}
  \item The KDC sends a message to the devices subscribed to by the leaving user to update their nonces as in Section~\ref{PP:3}: $n'_j = n_j + 1$. The new device key will be $DK'_j = h(ID_j \vert \vert n'_j)$. This helps in achieving forward secrecy.  
  \item The KDC sends the updated uKEKs to the SGs having  a common ancestor with SG\textsubscript{x}, encrypted by appropriate group keys or shared uKEKs. 
  \item For each DG subscribed to by the leaving user, the KDC sends a multicast containing the updated device keys to the  SGs subscribed to the DG, using the updated uKEKs. 
\end{itemize}

\subsection{When a New DG Joins}
\label{PP:8}
Recall that a SG is a group of users subscribing to a specific subset of DGs. 
So when a new DG joins, the possible number of SGs will change from $2^{P}-1$ to $2^{P+1}-1$, where $P$ is the number of DGs before the new DG joined, i.e., it will approximately double. Also, a given SG may be split into two unique SGs-- one with users with the same subscriptions as the original SG, and the other with users who opt to additionally subscribe to the new DG. 

\begin{figure}
  \centering
  \includegraphics[scale=0.29]{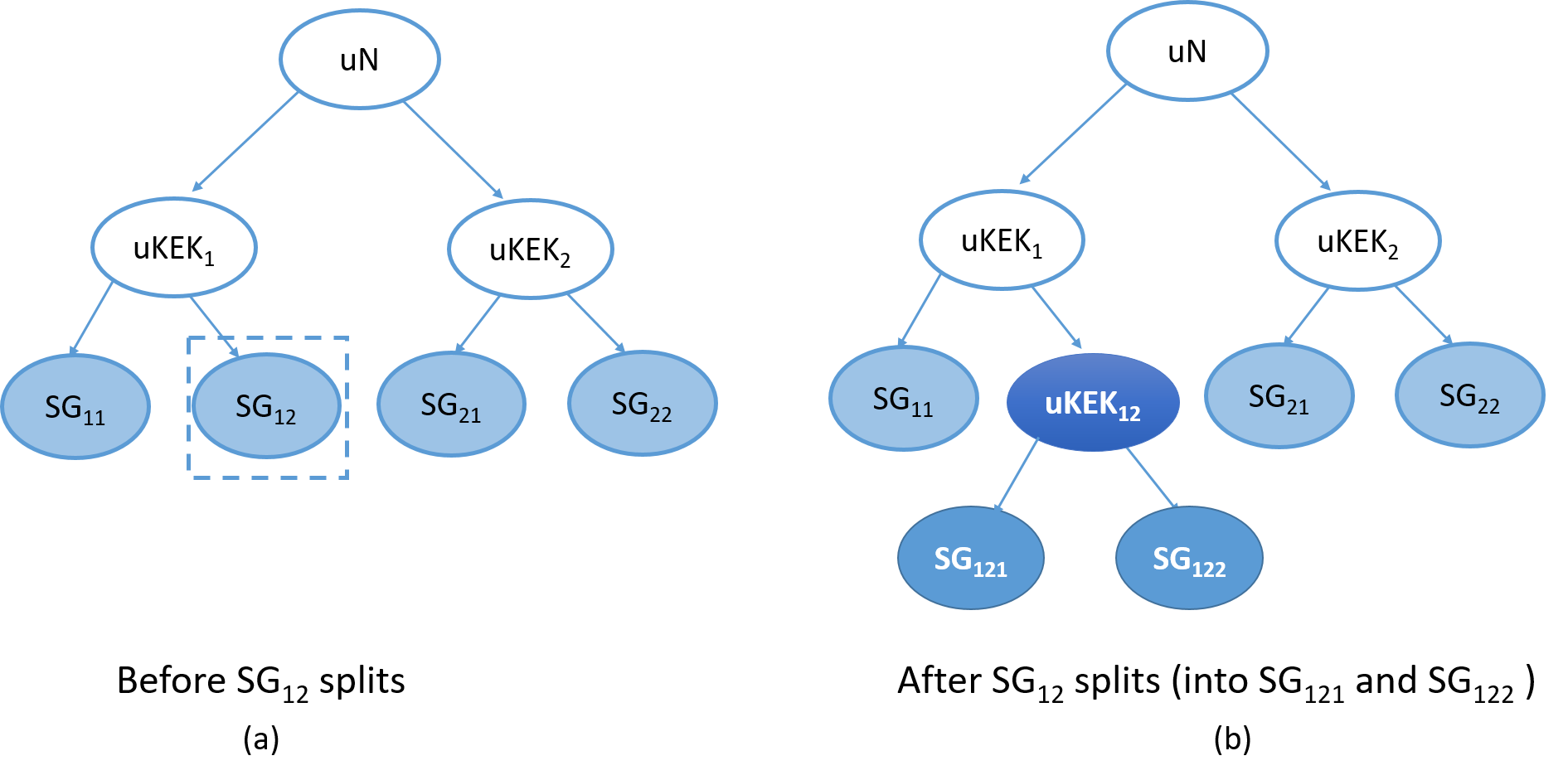}
  \caption{The figure on the left (respectively, right) shows the outer LKH tree before (respectively, after) SG\textsubscript{12} splits into two SGs-- SG\textsubscript{121} and SG\textsubscript{122}. The group key of SG\textsubscript{12} becomes the uKEK, uKEK\textsubscript{12}, for the users in the two SGs.}
  \label{fig12}
\end{figure}

This splitting is performed as follows: in the outer LKH tree, the two newly formed SGs become child nodes of the node corresponding to the original SG. In the example in Fig.~\ref{fig12}, when SG\textsubscript{12} splits, SG\textsubscript{121} and SG\textsubscript{122} become child nodes of the node corresponding to SG\textsubscript{12}. Also, the group key of SG\textsubscript{12} becomes the uKEK for both the child nodes after splitting; this uKEK is uKEK\textsubscript{12} in Fig.~\ref{fig12}. As a result, splitting does not necessitate updation of uKEKs to maintain forward or backward secrecy. However, we observe that the inner LKH trees  of the newly formed SGs need to be constructed appropriately based on which users from the original SG opt to subscribe to the new DG.  The KDC constructs these inner LKH trees, securely sends their KEKs and group keys to the newly formed SGs, and sends, to the newly formed SGs that subscribe to the newly joined DG,  the device keys of the devices of the DG. The splitting of a SG can cost heavy communication overhead to the network; however, since in practice, a new DG would only rarely join the network, this overhead would be small. We will later show that even under this rare event, the per user computation is small under our proposed protocol (see Sections~\ref{performance:B}.7 and~\ref{performance:C}.7). The following actions are performed:
\begin{itemize}
  \item For each SG that splits into two SGs, the KDC modifies the outer LKH tree by making the new SGs the child nodes of the original SG's node as explained above. 
  \item Whenever a SG is split into two SGs, the KDC constructs the inner LKH trees of the two SGs appropriately and securely distributes the group keys and KEKs of the newly formed SGs to the members of the SGs. 
  \item The KDC arranges the devices of the new DG in a LKH tree and generates its group key  and KEKs. 
  \item As in Section~\ref{PP:4}, the KDC establishes  a secret key, creates a secret identity and a nonce for each of the devices in the newly joined DG. 
  \item The KDC provides each of the devices in the newly joined DG with the respective individual (identity and nonce) and group  (group key and KEKs) secrets using the established shared secret keys between the devices and the KDC. Then the KDC as well as the devices derive the device keys. 
 \item Whenever a SG is split into two SGs, the KDC securely multicasts, to the members of the newly formed SG that subscribes to the newly joined DG,  the device keys of the devices of the DG, using the group key of the inner LKH tree of the SG. 
\end{itemize}

\subsection{When an Existing DG Leaves}
\label{PP:9}
\begin{figure}
  \centering
  \includegraphics[scale=0.29]{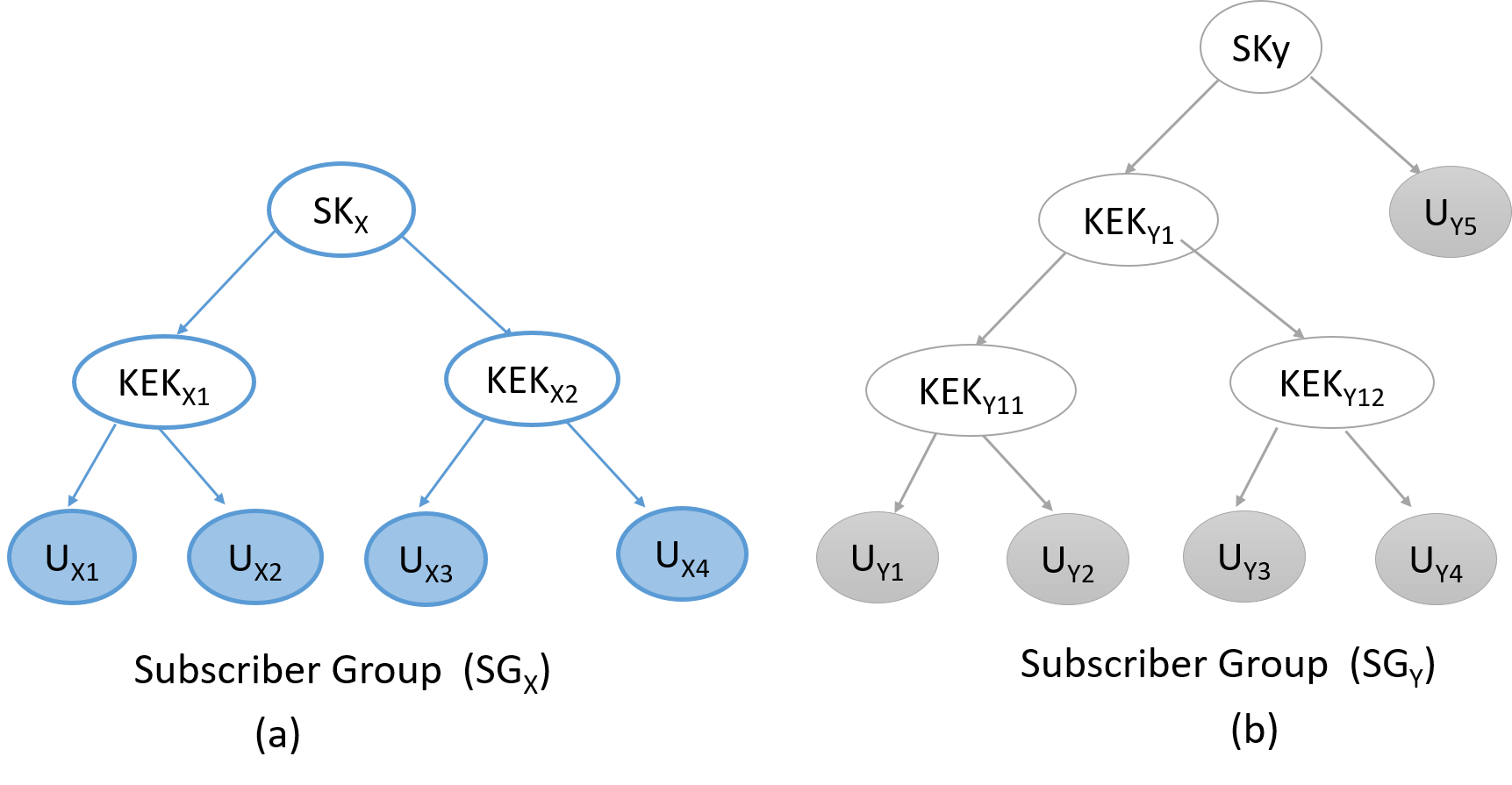}
  \caption{The figure shows the inner LKH trees of the groups SG\textsubscript{X} and SG\textsubscript{Y} before merging.}
  \label{fig13}
\end{figure}

\begin{figure}
  \centering
  \includegraphics[scale=0.31]{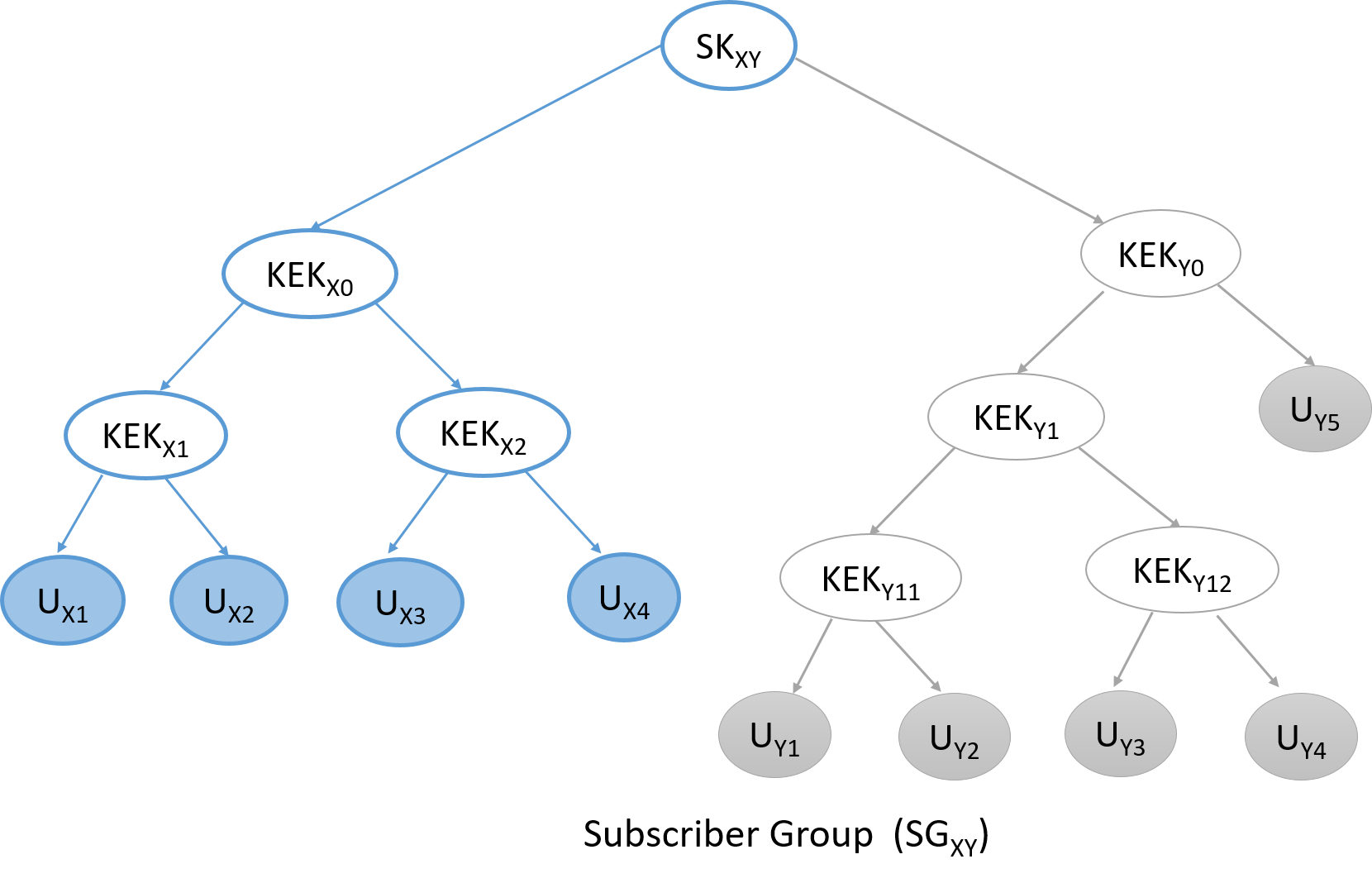}
  \caption{The figure shows the inner LKH tree of the group SG\textsubscript{XY} obtained by merging of the SGs 
SG\textsubscript{X} and SG\textsubscript{Y}. Note that SK\textsubscript{XY} is securely multicast to U\textsubscript{X1}, U\textsubscript{X2}, U\textsubscript{X3} and U\textsubscript{X4} by encrypting it using KEK\textsubscript{X0}.}
  \label{fig14}
\end{figure}

This case is exactly opposite to the case when a new DG joins (Section~\ref{PP:8}). In particular, recall that a SG is a group of users subscribing to a specific subset of DGs. 
So when an existing DG leaves, the possible number of SGs will change from $2^{P}-1$ to $2^{P-1}-1$, where $P$ is the number of DGs before the existing DG leaves, i.e., it will approximately halve.  Also, merging of pairs of SGs may be needed. Specifically, suppose before the DG left, there were two SGs, SG\textsubscript{X} and SG\textsubscript{Y}, which subscribed to the same set of DGs, except that SG\textsubscript{X} (respectively, SG\textsubscript{Y}) subscribed (respectively, did not subscribe) to the leaving DG. Then after the DG leaves, SG\textsubscript{X} and SG\textsubscript{Y} will be subscribed to the same set of DGs and hence need to be merged, forming a bigger SG, say SG\textsubscript{XY}. Under our proposed protocol, this merging is performed as follows. The inner LKH tree structures of SG\textsubscript{X} and SG\textsubscript{Y} are not changed; also, the root nodes (corresponding to group keys) of the two SGs become children (corresponding to KEKs) of the new root node of SG\textsubscript{XY}. Fig.~\ref{fig13} shows the inner LKH trees of two SGs, SG\textsubscript{X} and SG\textsubscript{Y}, before merging and Fig.~\ref{fig14} shows the inner LKH tree of the new SG, SG\textsubscript{XY}, formed after merging. The nodes (root nodes) corresponding to the group keys,  SK\textsubscript{X} and SK\textsubscript{Y}, of the two SGs become child nodes, corresponding to the KEKs KEK\textsubscript{X0} and KEK\textsubscript{Y0}, respectively, of the new root node corresponding to the group key SK\textsubscript{XY}. Also, SK\textsubscript{XY} can simply be the hash of the group key of the bigger group out of SG\textsubscript{X} and SG\textsubscript{Y}, say SG\textsubscript{Y}. Additionally, similar to Section~\ref{PP:7}, the LKH tree keys of the leaving DG need not be updated since new keys are derived if and when the DG starts re-populating. Also, the KDC sends a broadcast message to all SGs that the devices in the leaving DG are no longer available. For each instance of merging of two SGs, SG\textsubscript{X} and SG\textsubscript{Y}, into a bigger SG, SG\textsubscript{XY}, the following actions are performed:  
\begin{itemize}
  \item The KDC arranges the LKH trees of the two SGs as shown in Fig.~\ref{fig14}, and the KDC and users in the bigger SG, SG\textsubscript{Y},  calculate the new group key of the merged group SG\textsubscript{XY} as SK\textsubscript{XY} = h(SK\textsubscript{Y}). 
  \item The KDC securely multicasts SK\textsubscript{XY} to all the users of the subscriber group SG\textsubscript{X} by encrypting it using the key KEK\textsubscript{X0}. 
\end{itemize}
Also, unlike Section~\ref{PP:8}, modification of the outer LKH tree post-merging is, in general, non-trivial and depends on which pairs of SGs are merged. 
The KDC constructs the new outer LKH tree obtained after merging, and securely distributes its updated uKEKs to all the SGs using appropriate uKEKs or group keys. Although this may incur significant communication cost, since the leaving of an existing DG would typically be a rare event in practice, the resulting overhead would be small. Also, the per user computational cost incurred when an existing DG leaves is small (see Sections~\ref{performance:B}.8 and ~\ref{performance:C}.8).     

\subsection{Collusion}
\label{PP:10}
In~\cite{Kung}, the authors discussed about a possible collusion attack, which can occur under their proposed scheme, GroupIt, wherein a leaving user colludes with an existing device. This happens because the old user is aware of the device IDs of all the subscribed devices and the existing device knows the updated TEK. They can combine their knowledge to derive the other devices' new device keys. However, in our proposed scheme, users are not aware of any device's ID. Even if a user knows the device keys of all its subscribed devices, whenever a user leaves, the device keys are updated as explained in Section~\ref{PP:3} and Section~\ref{PP:7}. Hence, there is no possibility of a collusion attack under our proposed scheme. 

\subsection{Discussion}
Throughout, we have focused on designing a lightweight, flexible and secure GKM scheme. The incorporation of the outer LKH tree structure in the proposed scheme makes key distribution (especially that of device keys) across SGs highly efficient. Unlike GroupIt, which uses asymmetric encryption, the use of only symmetric encryption under our proposed scheme has also added to its efficiency. Also, we have provided increased flexibility by including the ability to handle the cases where a new DG joins, an old DG leaves, a new user occupies an empty SG, and an existing user leaves the last occupied spot in a SG in our proposed scheme; note that GroupIt does not explicitly handle any of the above events. We have also achieved better security as compared to GroupIt by maintaining forward and backward secrecy during all events of user and group dynamics, as well as completely preventing  collusion attacks.   


\section{Performance Analysis}
\label{performance}

\begin{table}
\centering
\caption{Notation used in Sections~\ref{performance} and~\ref{numerical}.}
\label{Index}
\begin{tabular}{|c|c|}
\hline
Notation & Meaning\\
\hline
$L_y$ &  No. of SGs subscribed to DG\textsubscript{y}  \\ 
$N_x$ & No. of users in  SG\textsubscript{x}  \\
$M_y$ & No. of devices in  DG\textsubscript{y}  \\
$M_{max}$ & Maximum no. of devices in any DG \\
$Y_x$ & No. of DGs subscribed to by  SG\textsubscript{x}  \\
$\lceil x \rceil$ & Ceiling of $x$ \\
$P$ &  No. of DGs  \\
$Q$  & No. of SGs \\
Dec & Symmetric Decryption \\
AsyDec & Asymmetric Decryption \\
Hash & Cryptographic hash function \\
 \hline
\end{tabular}
\end{table}
In Sections~\ref{performance:A},~\ref{performance:B}, and~\ref{performance:C}, we analyze the storage, communication, and computational costs, respectively, of our proposed protocol and compare them with those under the GroupIt protocol.   Table~\ref{Index} lists the notation used in this and the next section. 

\subsection{Storage}
\label{performance:A}
Under our proposed protocol, a user of group SG\textsubscript{x} stores at most $Y_x M_{max}$ device keys (the total number of DGs subscribed to by the user multiplied by the maximum number of devices in a DG, since there is a key for each device in each subscribed DG), at most $\lceil \log(N_x) \rceil$ KEKs (for multicast communication within SG\textsubscript{x}), one secret key (for message communication between the KDC and the user), one group key (for multicast messages between the KDC and the users of SG\textsubscript{x}), and at most $P$ uKEKs, which includes the key corresponding to the root uNode (for message communication using the outer LKH tree keys). On the other hand, a device of group DG\textsubscript{y} stores one ID and one nonce (for device key generation), at most $\lceil \log(M_y) \rceil$ KEKs (for multicast communication within DG\textsubscript{y}), one secret key (for message communication between the KDC and the device), and one group key (for multicast communication between the KDC and devices of DG\textsubscript{y}).

In GroupIt, a user of group SG\textsubscript{x} stores at most $Y_x M_{max}$ device IDs, $Y_x$ TEKs, at most $\lceil \log(N_x) \rceil$ KEKs, one secret key, and one group key. On the other hand, a device of group DG\textsubscript{y} stores one ID, at most $\lceil \log(M_y) \rceil$ KEKs, one secret key, one TEK, and one group key. 

We can see that in terms of storage, both our proposed protocol and GroupIt have a similar overhead for devices, and depending on the values of $P$ and $Y_x$, 
GroupIt may marginally outperform the proposed protocol with respect to the costs borne by the users.

\subsection{Communication}
\label{performance:B}
\subsubsection{New User Joining Group SG\textsubscript{x}}
Under our proposed protocol, the KDC establishes one secret key with the user, sends one broadcast message for the updation of device keys and the group key of SG\textsubscript{x}, at most $\lceil \log(N_x) \rceil$ multicasts to update KEKs, at most $P$ multicasts to update uKEKs and one unicast to share key material. Under GroupIt, the KDC establishes one secret key with the user, sends one broadcast message to update TEK, at most $\lceil \log(N_x) \rceil$ multicasts to update KEKs and one unicast to share key material. The  communication cost under our proposed protocol is higher than that under GroupIt; however, this is because of the provision of backward secrecy under our proposed protocol, but not under GroupIt (see Section~\ref{GroupIt:B}.3).  

\subsubsection{Existing User Leaving Group SG\textsubscript{x}}
Under our proposed protocol, the KDC sends one broadcast message to the devices subscribed to by the leaving user for the updation of device keys, at most $\lceil \log(N_x) \rceil$ multicasts to update KEKs and the group key, at most $P$ multicasts to update uKEKs and $Y\textsubscript{x}P$ multicasts for the updation of device keys with the remaining users. Under GroupIt, the KDC sends at most $\lceil \log(N_x) \rceil$ multicasts to update KEKs and the group key, at most $Y_x M_{max}$ multicasts and $Y_x$ broadcasts  for devices subscribed to by the leaving user and users subscribed to those devices to update their TEKs, respectively, and at most $Y_x \log(M_{max})$ multicasts to update device IDs. Hence, the proposed protocol is highly efficient in terms of communication overhead when an existing user leaves. This efficiency further improves if the number of devices within a DG increases or if the SG is subscribed to a large number of DGs. Through our solution, we have also eliminated the possibility of collusion, which exists under the GroupIt protocol (see Section~\ref{SSSC:GroupIt:B:collusion}).   

\subsubsection{New Device Joining Group DG\textsubscript{y}}
Under our proposed protocol, the KDC establishes one secret key with the device, sends one unicast containing the ID and nonce, one multicast for other members of the DG to update the group key, at most $\lceil \log(M_y) \rceil$ multicasts to update KEKs and at most $L_y$ multicasts for delivering the device's key to its subscribed users. Under GroupIt, the KDC establishes one secret key with the device, sends one multicast for other members of the DG to update the group key, at most $\lceil \log(M_y) \rceil$ multicasts to update KEKs, one broadcast containing the device's ID for the subscribed users and one unicast to share key material with the new device. We can see that if $L_y > 1$, then depending on the positions of the SGs subscribed to the DG of the new device in the outer LKH tree, GroupIt may have a lower communication cost in this case as compared to our proposed scheme.

\subsubsection{Existing Device Leaving Group DG\textsubscript{y}}
Both under our proposed protocol and GroupIt, the KDC sends one broadcast notifying about the exit of the device, and at most $\lceil \log(M_y) \rceil$ multicasts to update the KEKs and group key of the devices in DG\textsubscript{y}.  

\subsubsection{New User Occupying an Empty SG}
Under our proposed protocol, the KDC establishes one secret key with the user, sends one broadcast message for the updation of device keys with the users who have common subscriptions with the new user, at most $P$ multicasts to update uKEKs and one unicast to share key material with the new user. GroupIt does not explicitly handle this scenario.

\subsubsection{Existing User Leaves the Last Spot in Group SG\textsubscript{x}}
Under our proposed protocol, the KDC sends one broadcast message to the devices subscribed to by the leaving user for the updation of device keys, at most $P$ multicasts to update uKEKs and $Y\textsubscript{x}P$ multicasts for the updation of device keys with the  remaining users. Again, GroupIt does not explicitly handle this scenario.

\subsubsection{New Device Group DG\textsubscript{y} Joins}
We saw earlier that depending on which users subscribe to DG\textsubscript{y}, the communication cost for KEK updates may vary. In the worst case scenario, for a subscriber group SG\textsubscript{x}, of which every alternate member, going from left to right in its inner LKH tree, opts for this subscription, the KDC needs to send $N_x$ unicasts encrypted with the respective shared secret keys, one to each member of the two groups formed after SG\textsubscript{x} splits. The unicasts contain KEKs and group keys for the newly formed SG that does not subscribe to DG\textsubscript{y}, and they contain KEKs, group keys and the new devices' keys for the newly formed SG that subscribes to DG\textsubscript{y}. Additionally, under our proposed protocol, the KDC performs $M_y$ key establishments, one  with each of the devices of the new DG, and sends $M_y$ corresponding unicasts containing IDs, nonces, group key and KEKs. GroupIt does not explicitly handle this scenario.

\subsubsection{An Existing DG Leaves}
Under our proposed protocol, the KDC sends one broadcast notifying about the exit of the DG. For each instance of merging of two SGs, the KDC sends one multicast to the users of the bigger SG to hash their group key, and sends one multicast to the users of the smaller SG containing the new group key. We also observe that the number of uKEK updates required depends on the positions of different SGs in the outer LKH tree, and in the worst case scenario, the KDC sends one multicast to the members of each SG after merging, containing the updated uKEKs encrypted by the respective group keys. As before, GroupIt does not handle this scenario.

\subsubsection{Summary}
We observe that when a new user joins, our proposed protocol requires more multicasts than GroupIt to update uKEKs, which helps in achieving backward secrecy under our proposed protocol, which is not provided under GroupIt. 
However, GroupIt is significantly outperformed by our scheme when an existing user leaves, owing to the use of symmetric encryptions under our proposed protocol, unlike under GroupIt, which uses asymmetric encryption. In case of a device joining or leaving, both schemes have similar communication overhead.

\subsection{Computation}
\label{performance:C}
In this section, we analyze the computational costs per user and per device. 
\subsubsection{New User Joining Group SG\textsubscript{x}}
Under our proposed protocol, a device needs to perform one hash to update its device key. Each existing user performs one hash to update the group key, at most $Y_x M_{max}$ hashes to update device keys and two symmetric decryptions to update KEKs and uKEKs, whereas the new user performs one symmetric decryption to get the key material. Under GroupIt, each device needs to perform two hashes to update its device key. Each existing user performs $O(\log(N_x))$ symmetric decryptions to update KEKs, whereas the new user performs one symmetric decryption to get the key material. Hence, our proposed protocol outperforms GroupIt in terms of the computational cost for both devices and users. 

\subsubsection{Existing User Leaving Group SG\textsubscript{x}}
Under our proposed protocol, a device needs to perform one hash to derive the new device key. Each remaining user performs two symmetric decryptions to update KEKs and uKEKs, and $Y_x$ symmetric decryptions to get the updated device keys. Under GroupIt, a device needs to perform two symmetric decryptions and two hashes to derive the device keys. Each remaining user performs two symmetric and one asymmetric decryptions to gain KEKs and update information, and at most $Y_x M_{max}$ hashes to derive the new device keys. It is evident that our proposed protocol significantly improves upon GroupIt in terms of the overhead for devices, and also eliminates the need for asymmetric decryption, which is computationally expensive, for users.

\subsubsection{New Device Joining}
Under both our proposed protocol and GroupIt, the new device performs one symmetric decryption and one hash to derive its device key, whereas each existing device performs one hash to update the group key and one symmetric decryption to get the updated KEKs. Each subscribed user under our protocol performs one symmetric decryption to get the new device's device key, while it needs to perform an additional hash computation under GroupIt.

\subsubsection{Existing Device Leaving}
Under both our proposed protocol and GroupIt, each remaining device needs to perform one symmetric decryption to get the updated KEKs and group key, and users need not perform any computation.

\subsubsection{New User Occupying an Empty User Group SG\textsubscript{x}}
Under our proposed protocol, each device needs to perform one hash to update the  device key. Each user having common subscriptions with the new user performs $Y_x M_{max}$ hashes to update device keys and each user in a SG having a common ancestor with SG\textsubscript{x} performs one symmetric decryption to update uKEKs, whereas the new user performs one symmetric decryption to get the key material. GroupIt does not explicitly handle this scenario.

\subsubsection{Existing User Leaving the Last Spot in Group SG\textsubscript{x}}
Under our proposed protocol, each device needs to perform one hash to derive the new device key. Each user in an  SG having  a common ancestor with SG\textsubscript{x} performs one symmetric decryption to update uKEKs, and $Y_x$ symmetric decryptions to get the updated device keys. Again, GroupIt does not explicitly handle this scenario.

\subsubsection{New DG Joining}
Under our proposed protocol, each user that occupies a new leaf node in an inner LKH tree after splitting of an SG performs one symmetric decryption to get the updated KEKs, group keys and new devices' keys (if subscribed). Each  device in the new DG  needs to perform one symmetric decryption and one hash to get the key material and derive its device key. GroupIt does not handle this scenario.

\subsubsection{An Existing DG Leaving}
Under our proposed protocol, for every instance of merging of two SGs, each of the users in the bigger SG performs one hash, while each of the users in the smaller SG performs one symmetric decryption to get the new group key. A user needs to perform at most one symmetric decryption to get the updated uKEKs. As before, GroupIt does not handle this scenario.

\subsubsection{Summary}
Evidently, our proposed scheme outperforms GroupIt in every  user join or leave scenario. When a new user joins, our scheme requires only two symmetric decryptions per user, in contrast to GroupIt, for which this number of decryptions scales with $N_x$. When a user leaves, under our proposed protocol, devices  need not perform any decryptions, whereas they perform two decryptions each under GroupIt. Users also gain a significant advantage under our proposed protocol since asymmetric decryption is not required. Also, in case of a new device joining, our proposed protocol marginally outperforms GroupIt in terms of the computation costs borne by users. 


\section{Numerical Results}
\label{numerical}
In this section, via numerical computations, we compare the performance of our protocol with that of GroupIt for the events of user joining and leaving a group SG\textsubscript{x}, in terms of the communication cost (Section~\ref{numerical:A}) and computation cost (Section~\ref{numerical:B}).  
Suppose there are $P = 10$ DGs and $Q = 2^P - 1 = 1023$ SGs. Also, suppose each DG contains $M$ devices, each SG contains $N$ users, each SG is subscribed to $Y$ DGs and $L$ SGs are subscribed to each DG. Note that since $Q = 2^P - 1$, the number of SGs subscribed to each DG will be $2^{P-1}$; therefore, $L = 512$. 

In~\cite{Piedra}, the performances of Advanced Encryption Standard (AES), the  SHA-256 cryptographic hash function, and Elliptical Curve Cryptography (ECC) in commercial and research sensor nodes were compared. This study shows that ECC is much more computationally intensive as compared to AES, which is, in turn, more computionally intensive than SHA-256. As in~\cite{Kung}, for concreteness, we assume that hashing (Hash) using SHA-256 takes $T_0 = 460$ ns,
encryption (Enc) or decryption (Dec) using AES-256 of block size $64$ takes $T_1 = 800$ ns $=1.74 T_0$,  and asymmetric decryption (AsyDec) using ECC-224 takes $T_2 = 114000$ ns $=247.83 T_0$. Note that these absolute values are for a particular processor; however, the relative trends that we observe in the following numerical results hold for other processors as well.

\subsection{Communication Cost}
\label{numerical:A}
We now calculate the number of messages exchanged when a user leaves group SG\textsubscript{x}, as a function of the aforementioned variables. The following equations are derived from Section~\ref{performance:B}.2.

\begin{figure}
  \centering
  \includegraphics[scale=0.65]{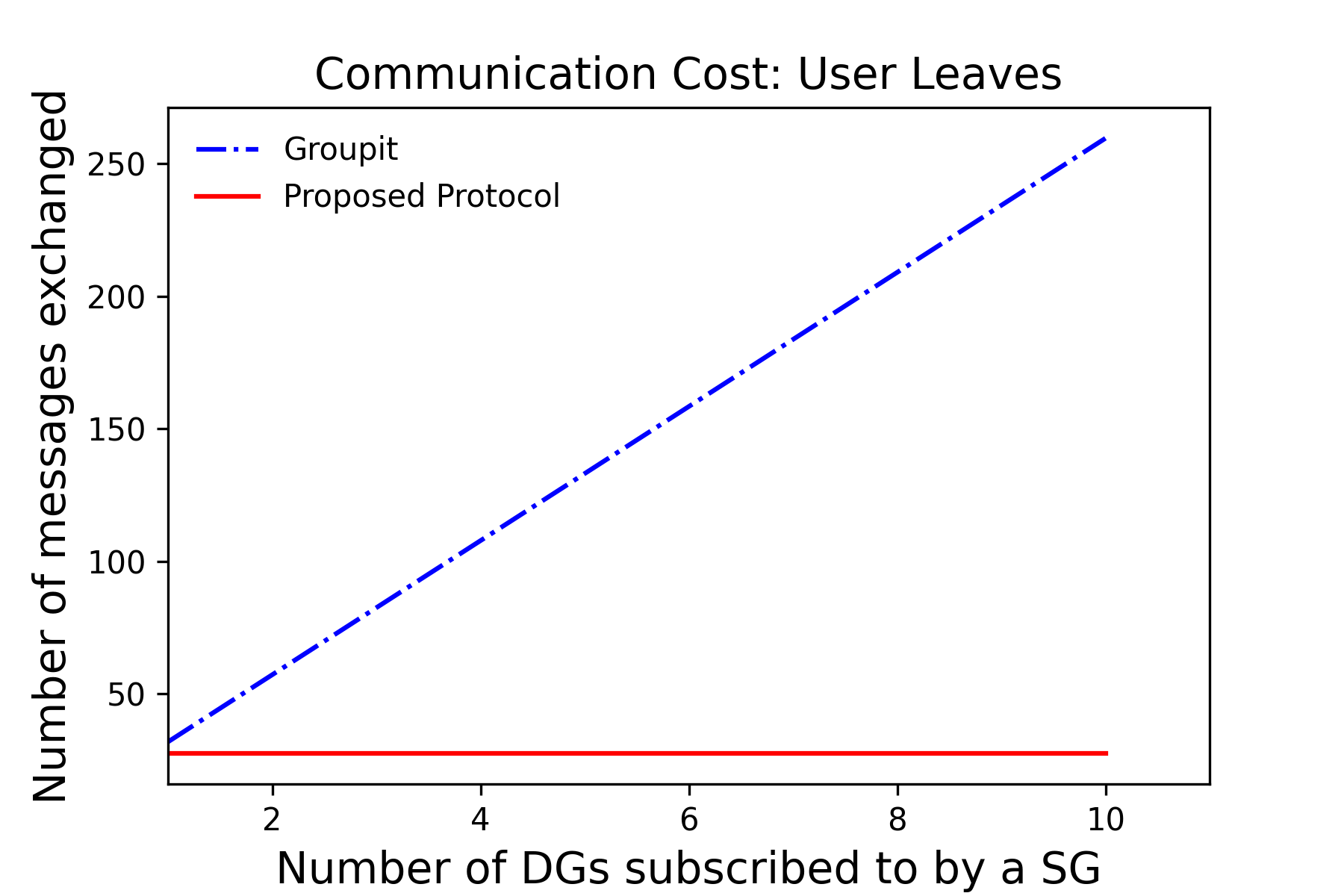}
  \caption{The figure shows the communication costs under GroupIt and the proposed protocol versus the number of DGs subscribed to by the user for the case when a user leaves group SG\textsubscript{x}.}
  \label{fig15}
\end{figure}

First, we fix $N$ = 100 and $M$ = 20, and vary the  
number of DGs subscribed to by the leaving user ($Y$). Under our proposed scheme, the number of messages exchanged is:
\begin{align*}
    1\mbox{ Broadcast} + (\lceil \log(N) \rceil + 2P)\mbox{ Multicasts}\\
    = 1\mbox{ Broadcast} + 27\mbox{ Multicasts}
\end{align*}
Under GroupIt, the number of messages exchanged is: 
\begin{align*}
    Y\mbox{ Broadcasts} + (\lceil \log(N) \rceil + Y\lceil \log(M) \rceil)\mbox{ Multicasts}\\ + M Y \mbox{ Unicasts}\\
    = Y \mbox{ Broadcasts} + (7 + 5 Y)\mbox{ Multicasts} + 20Y\mbox{ Unicasts} 
\end{align*}
The communication costs for both the protocols are depicted in Fig.~\ref{fig15}. Evidently, our scheme outperforms GroupIt as the number of messages required is low and does not depend on the value of $Y$. However, for GroupIt, it increases linearly in $Y$.

\begin{figure}
  \centering
  \includegraphics[scale=0.65]{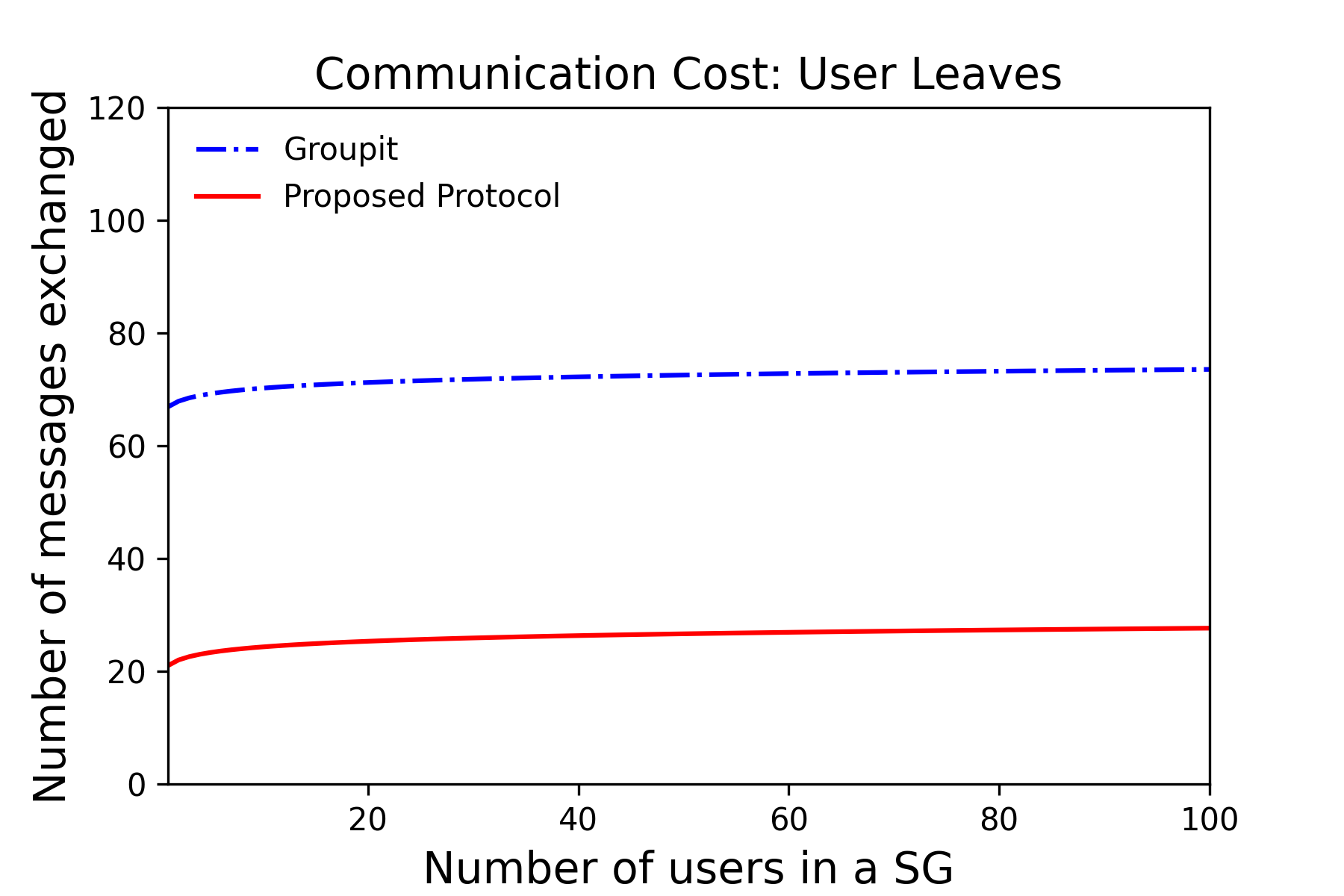}
  \caption{The figure shows the communication costs under GroupIt and the proposed protocol versus the number of users in group SG\textsubscript{x} for the case when a user leaves group SG\textsubscript{x}.}
  \label{fig16}
\end{figure}

Next, we fix $Y = 3$ and $M = 20$ and vary  $N$. The number of messages exchanged under our proposed scheme is: 
\begin{align*}
1\mbox{ Broadcast} + (\lceil \log(N) \rceil + 20)\mbox{ Multicasts}
\end{align*}
The number of messages exchanged under GroupIt is: 
\begin{align*}
3\mbox{ Broadcasts} + (\lceil \log(N) \rceil + 15)\mbox{ Multicasts} + 60\mbox{ Unicasts} 
\end{align*}
Again, as shown in Fig.~\ref{fig16}, our proposed protocol has lower communication cost than GroupIt. 

\begin{figure}
  \centering
  \includegraphics[scale=0.65]{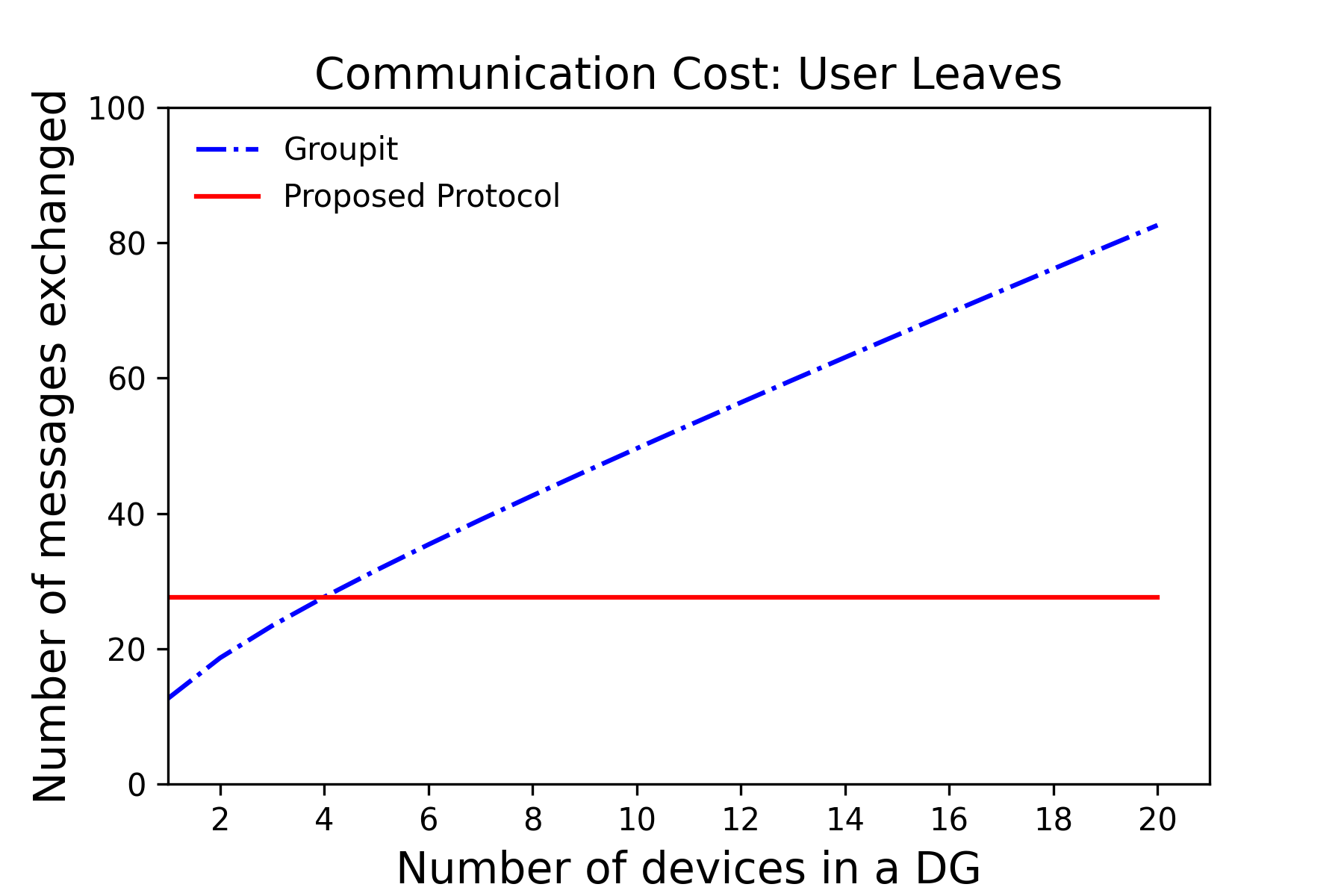}
  \caption{The figure shows the communication costs under GroupIt and the proposed protocol versus the number of devices in a DG for the case when a user leaves group SG\textsubscript{x}.}
  \label{fig17}
\end{figure}

Next, we fix $Y = 3$ and $N = 100$ and vary the number of devices  in each  DG ($M$). Under our proposed protocol, the number of messages exchanged is: 
\begin{align*}
1\mbox{ Broadcast} + 27\mbox{ Multicasts}. 
\end{align*}
On the other hand, under GroupIt, the number of messages exchanged is: 
\begin{align*}
3\mbox{ Broadcasts} + (7 + 3\lceil \log(M) \rceil)\mbox{ Multicasts} + 3M\mbox{ Unicasts}.  
\end{align*}
Fig.~\ref{fig17} shows that the communication cost under our proposed protocol does not increase in $M$, whereas that under GroupIt increases in $M$. 
The cost under our proposed protocol is lower than that under GroupIt when $M$ is sufficiently large. 

\subsection{Computation Cost}
\label{numerical:B}
We now study the computational costs borne by users and devices in the network under the proposed protocol and GroupIt when a new user joins and when an existing user leaves. We fix $Y$ = 3 in the rest of this section. The following computational costs are derived from Sections~\ref{performance:C}.1 and~\ref{performance:C}.2.

\begin{figure}
  \centering
  \includegraphics[scale=0.65]{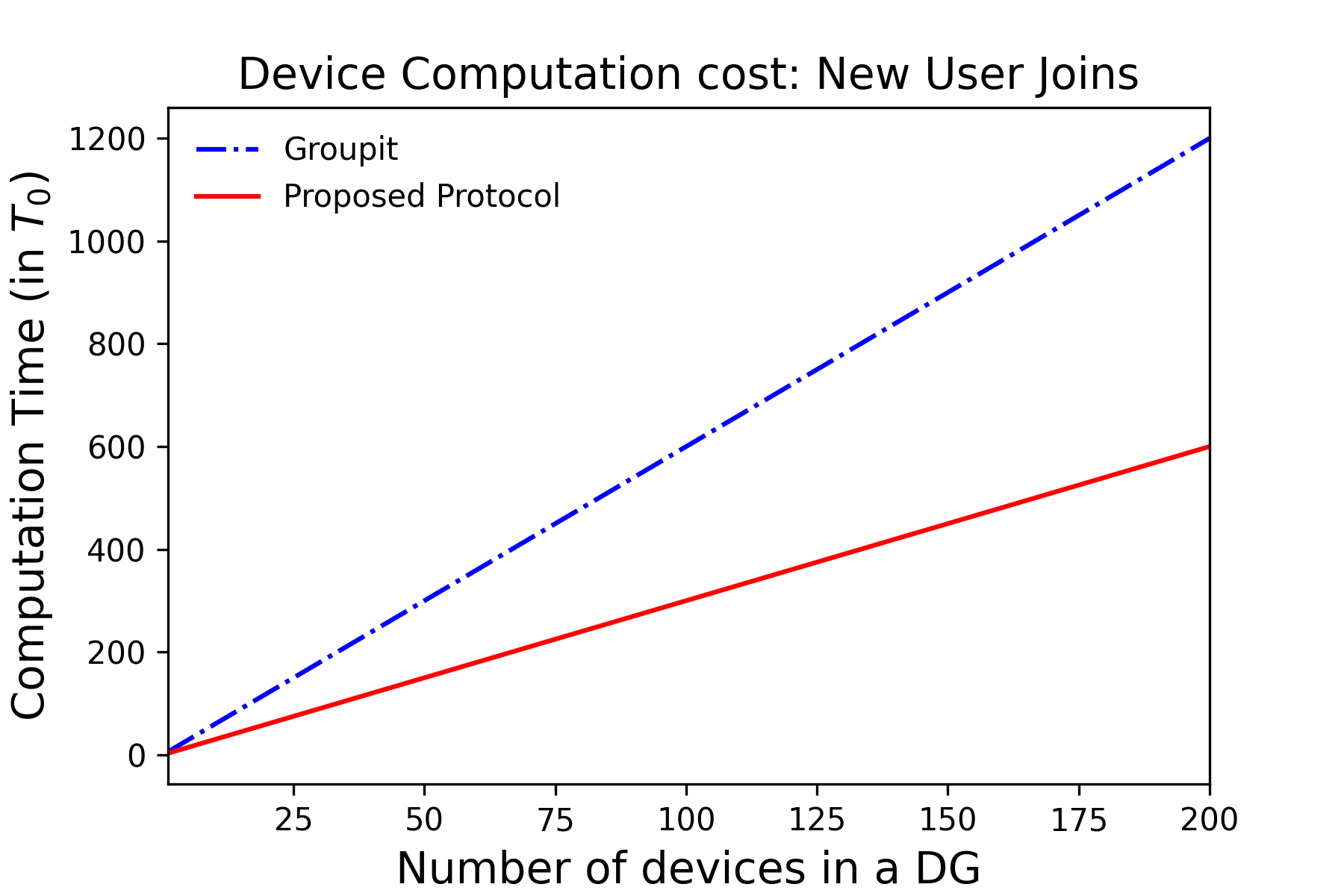}
  \caption{The figure shows the total computational load for all devices  under GroupIt and the proposed protocol versus the number of devices in a DG  for the case when a new user joins.}
  \label{fig18}
\end{figure}

First, when a new user joins SG\textsubscript{x}, the total computational load for all devices under the proposed protocol is: 
\begin{align*}
    M Y \mbox{ Hash} = 3T_{0}M
\end{align*}
For GroupIt, the load is: 
\begin{align*}
    2M Y\mbox{ Hash} = 6T_{0}M
\end{align*}
Both the above costs are plotted versus $M$ in Fig.~\ref{fig18}. Thus, when a new user joins SG\textsubscript{x}, the computational load for devices under our proposed protocol is half that under GroupIt.

\begin{figure}
  \centering
  \includegraphics[scale=0.6]{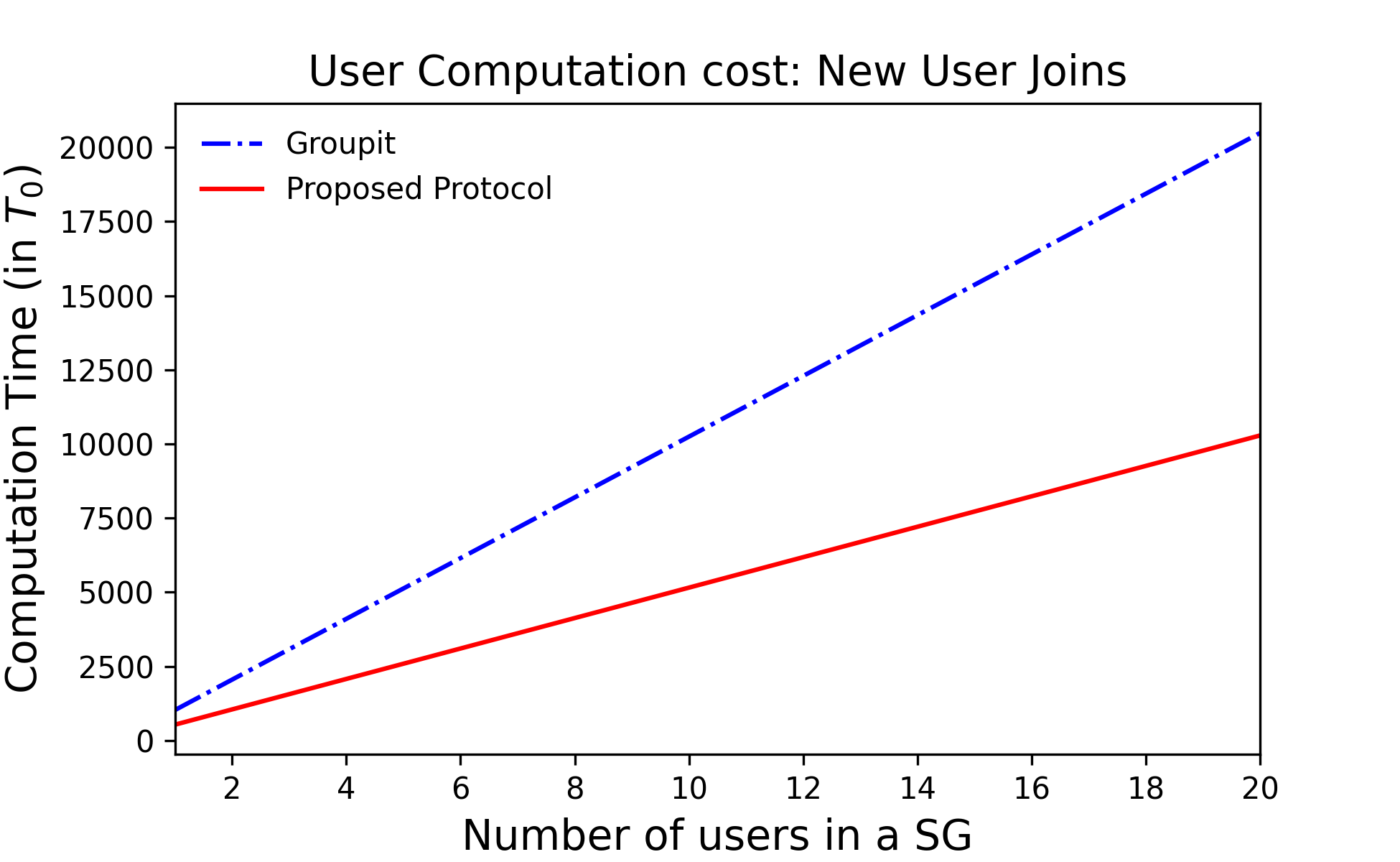}
  \caption{The figure shows the total computational load for all users  under GroupIt and the proposed protocol versus the number of users in a SG for the case when a new user joins.}
  \label{fig19}
\end{figure}

Now, when a new user joins SG\textsubscript{x}, the total computational load for all users under our proposed protocol is: 
\begin{align*}
  & &   L N \mbox{ Hash} + N \mbox{ Hash} + \log(N)\mbox{ Dec} + P\mbox{ Dec}\\
   & = & 513T_{0}N + 1.74T_{0} \log(N) + 17.4T_{0}\\
   & = & (513N + 1.74 \log(N) + 17.4)T_{0}
\end{align*}
For GroupIt, the corresponding load is: 
\begin{align*}
 & &   L N \mbox{ Hash} + L N \mbox{ Hash} + \log(N)\mbox{ Dec}\\
   &  = & 1024T_{0} N  + 1.74T_{0} \log(N)\\
   & = & (1024 N + 1.74 \log(N))T_{0}
\end{align*}
Fig.~\ref{fig19} shows the above costs versus the number of users in a SG ($N$).
We can see that for all, except very small values of $N$, our proposed protocol outperforms GroupIt. This advantage increases linearly in $N$.

\begin{figure}
  \centering
  \includegraphics[scale=0.65]{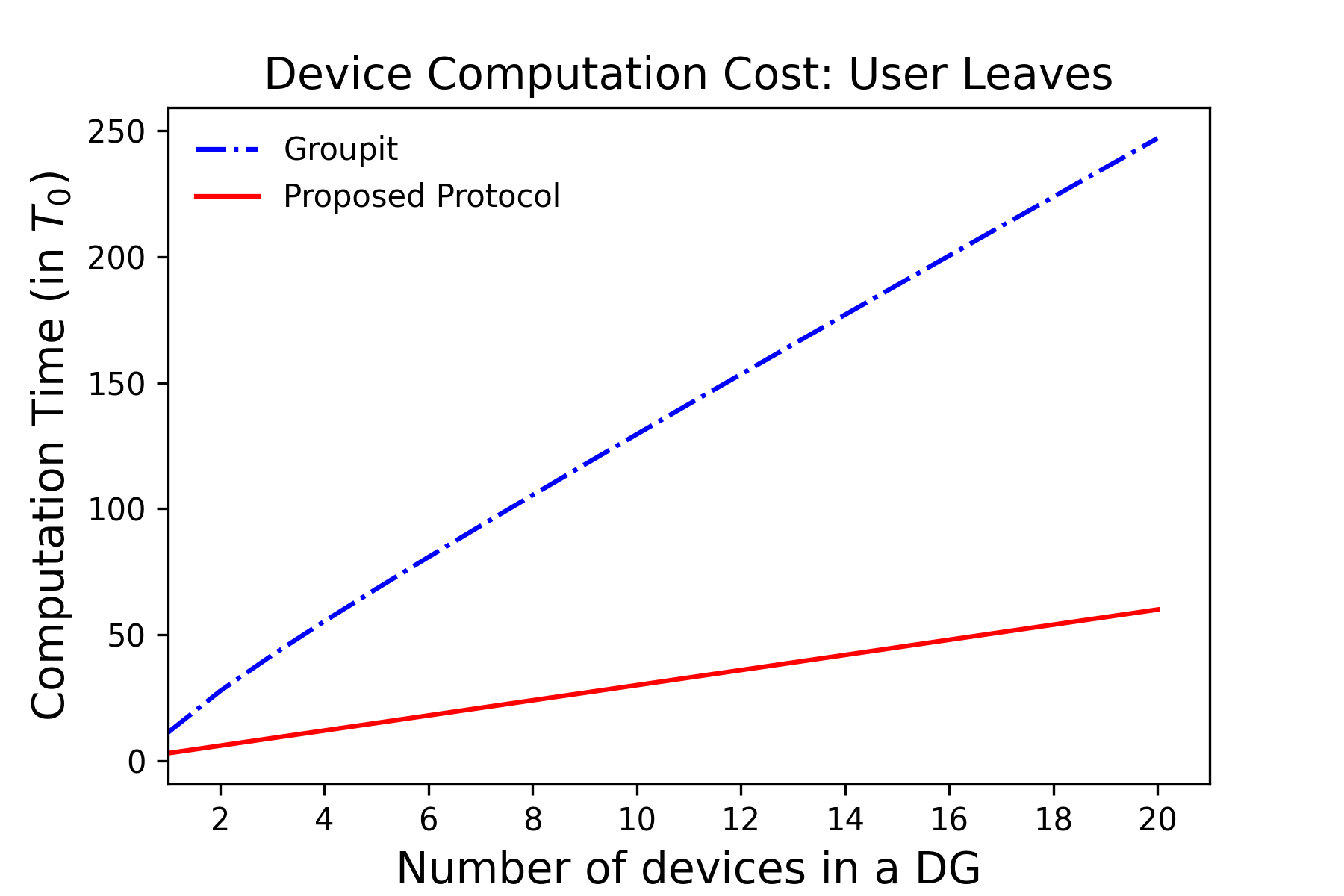}
  \caption{The figure shows the total computational load for all devices  under GroupIt and the proposed protocol versus the number of devices in a DG for the case when a user leaves SG\textsubscript{x}.}
  \label{fig20}
\end{figure}

Next, we examine the device computation cost when an existing user leaves SG\textsubscript{x}. Under our proposed protocol, the total computation cost for all devices is: 
\begin{align*}
    M Y \mbox{ Hash} = 3T_{0}M
\end{align*}
Under GroupIt, the corresponding cost is: 
\begin{align*}
& &    M Y \mbox{ Dec} + Y \log(M)\mbox{ Dec} + 2M Y \mbox{ Hash}\\
    & = & 5.22T_{0}M + 5.22T_{0} \log(M) + 6T_{0}M\\
   & = & (11.22M + 5.22 \log(M))T_{0}
\end{align*}
Fig.~\ref{fig20} shows that our proposed protocol significantly outperforms GroupIt in terms of the total computational load for all devices in the case when a user leaves SG\textsubscript{x}. This is due to the introduction of `nonces' as a part of the device key calculation in our proposed protocol, which eliminates the requirement to perform any decryption operation during device key updates.

\begin{figure}
  \centering
  \includegraphics[scale=0.65]{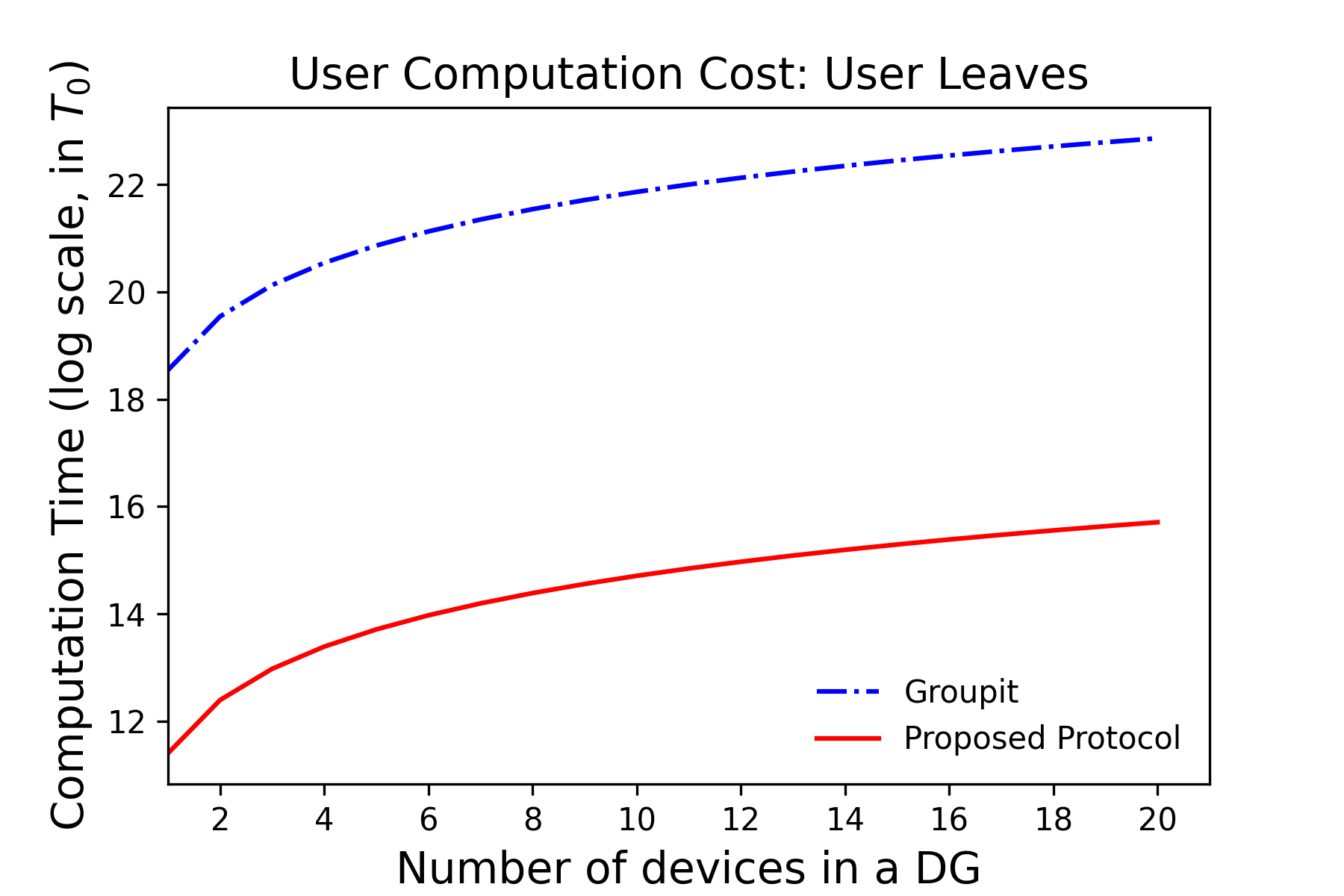}
  \caption{The figure shows the total computational load for all users  under GroupIt and the proposed protocol versus the number of devices in a DG for the case when a user leaves SG\textsubscript{x}.}
  \label{fig21}
\end{figure}

Finally, we consider the computational load incurred by users when an existing user leaves SG\textsubscript{x}. Under our proposed protocol, the total  computational cost incurred by all the users is: 
\begin{align*}
   & & \log(N)\mbox{ Dec} + \log(Q)\mbox{ Dec} + M Y L \mbox{ Dec}\\
    & = & 6.64\times{1.74T_{0}} + 10\times{1.74T_{0}} + 512\times{3}\times{1.74T_{0}M}\\
   & = & (28.95 + 2672.64 M)T_{0}
\end{align*}
The corresponding cost under GroupIt is: 
\begin{align*}
& &    \log(N)\mbox{ Dec} + M Y L \mbox{ Dec} + M Y L \mbox{ AsyDec}\\
   & = & 6.64\times{1.74T_{0}} + 1536\times{1.74T_{0}M} + 1536\times{247.83T_{0}M}\\
   & = & (11.55 + 383339.52M)T_{0}
\end{align*}
Fig.~\ref{fig21} shows the total user computation costs under both the schemes for the case when a user leaves SG\textsubscript{x} (note that the $y$-axis is on a log scale). The plot shows that our proposed protocol outperforms GroupIt by several orders of magnitude. This gain is achieved due to the use of symmetric encryptions under our proposed protocol in place of asymmetric double encryption, as well as due to the introduction of the outer LKH tree structure in our protocol.

\section{Conclusions and Future Work}
\label{conclusion}
We presented a highly efficient and secure GKM protocol for dynamic IoT
settings, which maintains forward and backward secrecy at all
times. Our proposed protocol uses only symmetric encryption, and is completely resistant to collusion attacks. Also, our protocol is highly flexible and can handle several new scenarios in which
device or user dynamics may take place, e.g., allowing a device
group to join or leave the network or creation or dissolution of a
user group, which are not handled by the GroupIt scheme. We evaluated the performance of the proposed protocol via extensive mathematical analysis and numerical computations, and showed that it outperforms the GroupIt scheme in terms of the communication and computation costs incurred by users and
devices. In this paper, we used the LKH scheme for GKM within SGs and DGs as well as by constructing an outer LKH tree structure, of which all the SGs are a part. One interesting direction for future work is to improve the performance of our proposed protcool by replacing LKH with other GKM schemes proposed in prior work such as OFT, ELK, MARKS etc.

\bibliographystyle{ieeetr}
\bibliography{references}

\begin{thebibliography}{10}

\bibitem{Maayan}
G.~Maayan, ``The iot rundown for 2020: Stats, risks, and solutions,'' {\em
  Security Today}, 13 Jan. 2020.

\bibitem{Das1}
A.~Das, A.~Sutrala, V.~Odelu, and A.~Goswami, ``A secure smartcard-based
  anonymous user authentication scheme for healthcare applications using
  wireless medical sensor networks,'' {\em Wireless Pers Commun}, vol.~94,
  pp.~1899--1933, 2017.

\bibitem{Jiang}
Q.~Jiang, S.~Zeadally, J.~Ma, and D.~He, ``Lightweight three-factor
  authentication and key agreement protocol for internet-integrated wireless
  sensor networks,'' {\em IEEE Access}, vol.~5, pp.~3376--3392, 2017.

\bibitem{Weber}
R.~Weber, ``Internet of things- new security and privacy challenges,'' {\em
  Computer Law \& Security Review}, vol.~26, pp.~23--30, 2010.

\bibitem{Granjal}
J.~Granjal, E.~Monteiro, and J.~S. Silva, ``Security for the internet of
  things: A survey of existing protocols and open research issues,'' {\em IEEE
  Communications Surveys $\&$ Tutorials}, vol.~17, no.~3, pp.~1294--1312, 2015.

\bibitem{Veltri}
L.~Veltri, S.~Cirani, S.~Busanelli, and G.~Ferrari, ``A novel batch-based group
  key management protocol applied to the internet of things,'' {\em Ad Hoc
  Networks}, vol.~11, pp.~2724--2737, 2013.

\bibitem{Rafaeli}
S.~Rafaeli and D.~Hutchison, ``A survey of key management for secure group
  communication,'' {\em ACM Comput. Surv.}, vol.~35, no.~3, p.~309–329, 2003.

\bibitem{Omar}
O.~Cheikhrouhou, A.~Koubâa, G.~Dini, H.~Alzaid, and M.~Abid, ``Lnt: A logical
  neighbor tree secure group communication scheme for wireless sensor
  networks,'' {\em Ad Hoc Networks}, vol.~10, pp.~1419--1444, 2012.

\bibitem{Esposito}
C.~Esposito, M.~Ficco, A.~Castiglione, F.~Palmieri, and A.~D. Santis,
  ``Distributed group key management for event notification confidentiality
  among sensors,'' {\em IEEE Transactions on Dependable and Secure Computing},
  vol.~17, no.~3, pp.~566--580, 2018.

\bibitem{Kung}
Y.~Kung and H.~Hsiao, ``Groupit: Lightweight group key management for dynamic
  iot environments,'' {\em IEEE Internet of Things Journal}, vol.~5, no.~6,
  pp.~5155--5165, 2018.

\bibitem{Lin}
J.~Lin, W.~Yu, N.~Zhang, X.~Yang, H.~Zhang, and W.~Zhao, ``A survey on internet
  of things: Architecture, enabling technologies, security and privacy, and
  applications,'' {\em IEEE Internet of Things Journal}, vol.~4, no.~5,
  pp.~1125--1142, 2017.

\bibitem{Sicari}
S.~Sicari, A.~Rizzardi, L.~Grieco, and A.~Coen-Porisini, ``Security, privacy
  and trust in internet of things: The road ahead,'' {\em Computer Networks},
  vol.~76, pp.~146--164, 2015.

\bibitem{Gutierrez}
J.~Gutierrez, M.~Naeve, E.~Callaway, M.~Bourgeois, V.~Mitter, and B.~Heile,
  ``{IEEE} 802.15.4: a developing standard for low-power low-cost wireless
  personal area networks,'' {\em IEEE Network}, vol.~15, no.~5, pp.~12--19,
  2001.

\bibitem{Sheng}
Z.~Sheng, S.~Yang, Y.~Yu, A.~Vasilakos, J.~Mccann, and K.~Leung, ``A survey on
  the {IETF} protocol suite for the internet of things: standards, challenges,
  and opportunities,'' {\em IEEE Wireless Communications}, vol.~20, no.~6,
  pp.~91--98, 2013.

\bibitem{Schrickte}
L.~Schrickte, C.~Montez, R.~Oliveira, and A.~Pinto, ``Integration of wireless
  sensor networks to the internet of things using a {6LoWPAN} gateway,'' in
  {\em 2013 III Brazilian Symposium on Computing Systems Engineering},
  (Niteroi), pp.~119--124, 2013.

\bibitem{Yang}
Y.~Yang, L.~Wu, G.~Yin, L.~Li, and H.~Zhao, ``A survey on security and privacy
  issues in internet-of-things,'' {\em IEEE Internet of Things Journal},
  vol.~4, no.~5, pp.~1250--1258, 2017.

\bibitem{Turkanovic}
M.~Turkanovic, B.~Brumen, and M.~Hölbl, ``A novel user authentication and key
  agreement scheme for heterogeneous ad hoc wireless sensor networks, based on
  the internet of things notion,'' {\em Ad Hoc Networks}, vol.~20, pp.~96--112,
  2014.

\bibitem{Porambage}
P.~Porambage, A.~Braeken, C.~Schmitt, A.~Gurtov, M.~Ylianttila, and B.~Stiller,
  ``Group key establishment for secure multicasting in iot-enabled wireless
  sensor networks,'' in {\em 2015 IEEE 40th Conference on Local Computer
  Networks (LCN)}, (Clearwater Beach, FL), pp.~482--485, 2015.

\bibitem{Das}
A.~Das, S.~Kumari, V.~Odelu, X.~Li, F.~Wu, and X.~Huang, ``Provably secure user
  authentication and key agreement scheme for wireless sensor networks,'' {\em
  Security and Communication Networks}, pp.~3670--3687, 2016.

\bibitem{Chang}
C.~Chang and H.~Le, ``A provably secure, efficient, and flexible authentication
  scheme for ad hoc wireless sensor networks,'' {\em IEEE Transactions on
  Wireless Communications}, vol.~15, no.~1, pp.~357--366, 2016.

\bibitem{Ferrari}
N.~Ferrari, T.~Gebremichael, U.~Jennehag, and M.~Gidlund, ``Lightweight
  group-key establishment protocol for iot devices: Implementation and
  performance analyses,'' in {\em 2018 Fifth International Conference on
  Internet of Things: Systems, Management and Security}, (Valencia),
  pp.~31--37, 2018.

\bibitem{Park}
C.~Park, ``A secure and efficient ecqv implicit certificate issuance protocol
  for the internet of things applications,'' {\em IEEE Sensors Journal},
  vol.~17, no.~7, pp.~2215--2223, 2017.

\bibitem{Eldefrawy}
M.~Eldefrawy, N.~Pereira, and M.~Gidlund, ``Key distribution protocol for
  industrial internet of things without implicit certificates,'' {\em IEEE
  Internet of Things Journal}, vol.~6, no.~1, pp.~906--917, 2019.

\bibitem{Robinson}
A.~Robinson and R.~Steinwandt, ``Group key establishment with physical
  unclonable functions,'' {\em Journal of Information and Optimization
  Sciences}, vol.~40, pp.~69--80, 2019.

\bibitem{Ferrari1}
N.~Ferrari, ``Context-based authentication and lightweight group key
  establishment protocol for iot devices,'' Master's thesis, Mid Sweden
  University, Faculty of Science, Technology and Media, Department of
  Information Systems and Technology, 2019.

\bibitem{Wong}
C.~Wong, M.~Gouda, and S.~Lam, ``Secure group communications using key
  graphs,'' {\em SIGCOMM Comput. Commun. Rev.}, vol.~28, no.~4, p.~68–79,
  1998.

\bibitem{Son}
J.~Son, J.~Lee, and S.~Seo, ``Topological key hierarchy for energy-efficient
  group key management in wireless sensor networks,'' {\em Wireless Pers
  Commun}, vol.~52, no.~359, 2009.

\bibitem{Mughal}
M.~Mughal, P.~Shi, A.~Ullah, K.~Mahmood, M.~Abid, and X.~Luo, ``Logical tree
  based secure rekeying management for smart devices groups in iot enabled
  wsn,'' {\em IEEE Access}, vol.~7, pp.~76699--76711, 2019.

\bibitem{Conti}
M.~Conti, R.~Pietro, L.~Mancini, and A.~Mei, ``Emergent properties: detection
  of the node-capture attack in mobile wireless sensor networks,'' in {\em
  Proceedings of the First ACM Conference on Wireless Network Security, WISEC
  2008}, (Alexandria, VA, USA), 2008.

\bibitem{Bonaci}
T.~Bonaci, L.~Bushnell, and R.~Poovendran, ``Node capture attacks in wireless
  sensor networks: A system theoretic approach,'' in {\em 49th IEEE Conference
  on Decision and Control (CDC)}, (Atlanta, GA), pp.~6765--6772, 2010.

\bibitem{Wang}
X.~Wang, J.~Zhang, E.~Schooler, and M.~Ion, ``Performance evaluation of
  attribute-based encryption: Toward data privacy in the iot,'' in {\em 2014
  IEEE International Conference on Communications (ICC)}, (Sydney, NSW),
  pp.~725--730, 2014.

\bibitem{Piedra}
A.~de~la Piedra, A.~Braeken, and A.~Touhafi, ``A performance comparison study
  of ecc and aes in commercial and research sensor nodes,'' in {\em Eurocon
  2013}, (Zagreb), pp.~347--354, 2013.

\end{thebibliography}

\end{document}